\documentclass[aps,prxquantum,reprint,onecolumn,superscriptaddress,longbibliography]{revtex4-2}
\pdfoutput=1
\usepackage{amssymb}
\usepackage{amsthm}
\usepackage{mathtools}
\usepackage{physics}
\usepackage{todonotes}
\usepackage{comment}
\usepackage{hyperref}
\usepackage{tikz}
\usepackage{float}
\usetikzlibrary{decorations.pathmorphing, backgrounds, positioning}

\begin{document}

\title{Non-abelian quantum cellular automata: $1{+}1$-dimensional $SU(2)$ Yang--Mills with fermions}

\author{Dogukan Bakircioglu}
\affiliation{Université Paris-Saclay, Inria, CNRS, LMF, 91190 Gif-sur-Yvette, France}
\email{dogukan.bakircioglu@inria.fr}

\author{Pablo Arrighi}
\affiliation{Université Paris-Saclay, Inria, CNRS, LMF, 91190 Gif-sur-Yvette, France}
\email{pablo.arrighi@inria.fr}

\begin{abstract}
This work provides a digital quantum simulation scheme for $1{+}1$-dimensional $SU(2)$ Yang--Mills theory with Dirac fermions. It takes the form of a quantum circuit, infinitely repeating across space and time with $\Delta_t=\Delta_x=\varepsilon$, whose wires follow lightlike propagation. The construction mirrors the logic of the standard quantum field theory approach, transposed to the discrete setting. Namely, we start from the Dirac quantum walk, restore $SU(2)$ gauge symmetry by introducing the gauge field, lift the walk to a multi-particle QCA while preserving fermionic anticommutation, and equip the gauge field with its own dynamics. Rather than relying on Clebsch--Gordan decompositions, we use the pointwise-product structure of gauge-link updates, which yields self-contained proofs of unitarity and gauge covariance in quantum-computing notation. The construction provides an explicit algorithmic formulation of the theory, whose continuum limit we discuss.
\end{abstract}
\maketitle

\section{Introduction}\label{sec:introduction}

\noindent {\em Formalisms.} Quantum field theories (QFT) describe interacting particles in the relativistic and quantum regimes. Their specifics are usually motivated by requiring commutation with the representations of a given gauge transformation group, which may or may not be abelian. QFTs are notoriously hard to formulate and to simulate, especially in the non-abelian case. In particular, they suffer from divergences, which must be tamed by regularisation/discretisation. Numerous works acknowledge this fact and take the view that the right approach is to formulate the theory in a natively discrete fashion, and then take a continuum limit. There are mainly two such natively discrete formulations of QFT, to which this work aims to add a third.

The most established way of formulating a discretized QFT is through a discrete Lagrangian---a scalar density on a spacetime lattice with $\Delta t = \Delta x = \varepsilon$, whose integral defines the action entering the Feynman path integral \cite{Wilson:1974,KogutSusskind1975}.
Its Feynman path integral-based dynamics can be evaluated numerically by means of Wick rotations and Monte Carlo methods. These numerical methods have two disadvantages, however: the Lorentzian symmetry is lost through switching to Euclidean space \cite{Smit2002,rothe,montvay1994quantum}, and the sign problem exhausts the capacities of classical computers \cite{Troyer:2004ge,Gattringer:2016kko}. Moreover, they do not immediately translate into a quantum computer algorithm.

The second most established way of formulating a discretized QFT is through its Hamiltonian---a discrete-space local operator generates the continuous-time dynamics. Through numerical techniques, e.g.\ deploying tensor network ansatzes \cite{Verstraete_2010,Shachar_2022,bañuls2013matrixproductstateslattice,Magnifico_2021,Magnifico_2025}, it becomes possible to evaluate the low-energy states of the theory, which are the most likely to occur in nature. Computing the evolution is again of exponential cost on classical computers. But the main advantage of this formulation is its quantum-computing compatibility: the evolution can be mimicked by that of a quantum device whose Hamiltonian dynamics can be tweaked into being similar, i.e.\ emulating one quantum physical system by another \cite{Martinez_2016}. This approach is called analog quantum simulation. Still, it has the disadvantage of being short-lived, because a fundamental assumption of this formalism is $\Delta t \ll \Delta x$. More generally, this asymmetric treatment of time versus space is the main disadvantage of this formalism, as it means that it fundamentally departs from the relativistic regime and breaks its symmetries.

We promote a third formalism, which seeks to gather the best of both worlds through drawing the lessons of the quantum computing tidal wave. The point is to treat space and time on an equal footing $\Delta t = \Delta x = \varepsilon$, and yet present the dynamics explicitly as a quantum simulation algorithm, in the spirit of constructive mathematics. In fact the theory is simply presented as a quantum circuit, infinitely repeating across discrete spacetime, and whose wires match the maximal speed of propagation of information. This approach is called digital quantum simulation. It has the advantage of being long-lived, because quantum error correction can be deployed throughout, and $\Delta t \not \ll \Delta x$. As regards symmetries, most can be treated at the level of individual quantum gates: unitarity is that of each gate, gauge covariance boils down to them commuting with the gauge transformation group. Lorentz covariance can either be treated natively in the discrete \cite{Arrighi-Facchini,PaviaLORENTZ,DebbaschLORENTZ} or postponed till the continuum limit.

\noindent {\em Quantum cellular automata.} Each step of the evolution is the application of a translation-invariant finite-depth quantum circuit of local quantum gates, namely a quantum cellular automaton (QCA) \cite{arrighi2019overviewquantumcellularautomata,Farrelly_2020}. This is also the architecture that Feynman envisioned when he invented the idea of ``quantum computer'' \cite{Feynman1982}. His motivations were the same; the main application for him was to efficiently simulate quantum physics. Much progress has been made since. In the one-particle sector, QCAs specialize into Quantum Walks (QW). We have for instance learned that those translation-invariant QW which have a continuous limit yield Dirac-like dynamics \cite{Pavia-Dirac}; that this can be made into a rigorous convergence result including in $3{+}1$ dimensions \cite{ArrighiNesmeForets2014}; that relaxing homogeneity allows us to model first-quantized electromagnetic fields \cite{Bisio_2016,arnault2015landaulevelsdiscretetimequantum} and further relaxing to two-step continuum limits yields the Dirac equation in curved spacetime \cite{debbasch2022minimalquantumwalksimulation,giuseppe2013quantumwalksmasslessdirac,arrighi2016quantumwalkingcurvedspacetime}. A well-known problem when formulating fermions on a lattice is fermion doubling; for the quantum walk models of interest here this issue is essentially settled \cite{bakircioglu2026fermiondoublingquantumcellular,gupta2026fermiondoublingdiracquantum,Jolly_2023}.

In the multi-particle sector of QCA, we have been stuck with structural \cite{arrighi2009unitaritypluscausalityimplies} and universality \cite{Watrous,ArrighiQGOL} results for some time, until a series of recent breakthroughs. First, Thirring-model-like fermion-fermion interactions were analysed \cite{Bisio_2018}. Then, we were able to express $1{+}1$ quantum electrodynamics (QED) \cite{ArrighiQED} and $3{+}1$ QED \cite{Eon_2023} as QCA, validating the limits of these towards the corresponding Kogut--Susskind Hamiltonian in \cite{sellapillay2022discreterelativisticspacetimeformalism}, thus establishing the ability to model fermion-boson interactions. Additional physical insights into QCAs have emerged from studying the Dirac sea \cite{Gupta_2025} and renormalization group properties \cite{Trezzini_2025}. The next challenge was tackling a non-abelian gauge theory; this is the main result of this paper.

\noindent {\em Challenges and results.} This paper provides the first QCA for an $SU(2)$-gauge Yang--Mills theory coupled to Dirac fermions. We do so in $1{+}1$ dimensions. We faced several major challenges in the process: (i)~The language that standard QFT books use \cite{Peskin:1995ev,Zee2010} is quite remote from that of quantum computing---much time was spent in translating and developing appropriate notations. (ii)~These books heavily rely on Clebsch--Gordan sums for their key steps, whose theory is hard to present in a self-contained way, and whose unitarity holds only in a very specific sense, insufficient for the sake of this paper. (iii)~This literature is generally inexplicit in comparison with modern standards in quantum computing---much effort was required in order to obtain precise and rigorous results. Our $1{+}1$ $SU(2)$ QCA adopts modern quantum computing notation. It does not use Clebsch--Gordan theory until the very end, in order to characterize the gauge-invariant subspace, when unitarity is not required. Instead, we rely on the pointwise-product structure of the gauge link updates. This choice allows for self-contained, terse, line-by-line, rigorous proofs of unitarity and gauge covariance. The end result is an explicit quantum circuit description of a quantum simulation algorithm for $1{+}1$ $SU(2)$-gauge Yang--Mills theory coupled to Dirac fermions. It may serve as a numerical scheme, whose continuum limit we discuss. But it may also serve as an alternative presentation of the theory itself.

\noindent {\em Plan.} In Sec.~\ref{sec:fermions} we set the scene and notations. Naive fermion hopping, from site $x$ to $x+1$, does not commute with gauge transformations, and the main issue will be to fix this. This motivates the introduction of the gauge field in Sec.~\ref{sec:gaugefield}, we explain exactly how gauge transformations act in Sec.~\ref{sec:gaugetransformations}, and finally obtain the commutation in Sec.~\ref{sec:covariance}. This however involves a controlled pointwise-product whose unitarity we check in Sec.~\ref{sec:unitarity}. The paper at this stage handles only the one-fermion sector. Sec.~\ref{sec:multi-particle} is a small self-contained presentation of previously-known-but-scattered techniques to extend any gate to the multiple-particle regime, whilst preserving unitarity and gauge covariance. The gauge field dynamics is turned on in Sec.~\ref{sec:interaction}, suggesting possible ways of truncating the theory. The gauge-invariant subspace and its corresponding Gauss law are given in Sec.~\ref{sec:gauss-law}. All the pieces are fitted together in Sec.~\ref{sec:altogether}, and the continuum limit to the Hamiltonian of $1{+}1$-dimensional $SU(2)$ Yang--Mills coupled to Dirac fermions is discussed. We summarise our results and describe future work in Sec.~\ref{sec:conclusion}.

\section{Fermions}\label{sec:fermions}

On each site $x$, some fermions are right-movers (annihilator operator $a_x$) and some fermions are left-movers (annihilator operator $b_x$); this degree of freedom is called `chirality'. These names are those of the transfer phase: it is the $a$ fermions that the transfer carries rightwards, from $x$ to $x+1$. Over a whole timestep the transfer is preceded by an on-site swap, which reverses the reading; we come back to this in Sec.~\ref{sec:unitarity}.
We focus on the transfer of fermions from a specific site $x$ to $x+1$. In the one-particle sector this will involve terms of the form $b^\dagger_{x+1} a_x$. Since there are no $a_{x+1}$ nor $b_{x}$ involved, we drop the site index and write $b^\dagger a$ instead.

We want the hopping to be $SU(2)$-gauge-covariant. The first non-trivial representation of the $SU(2)$ group is the $2\times 2$ special unitary matrices. Hence, for each chirality, there needs to be at least a $2$-dimensional `colour' degree of freedom upon which to act. It follows that our operators will in fact be $a^\pm$ and $b^\pm$.

Bearing in mind that fermions may change colour as they hop, the one-particle sector of the fermionic transfer will typically involve terms of the form

\begin{align}
T_R=\sum_{m,n} {b^n}^\dagger M^{*}_{mn}\, a^m .\label{eq:TR}
\end{align}

Consider
\begin{align}
G=\bigotimes_x G_x \label{eq:G}
\end{align}
an $SU(2)$-gauge transformation. In Sec.~\ref{sec:gaugetransformations} we will describe precisely how each $G_x$ acts, but we can already safely say that
\begin{align}
    G_x(a^m) := G_x a^m G_x^\dagger = a^m G_x^\dagger = \sum_{m'} a^{m'}\, (g_x^\dagger)_{mm'} \label{eq:atransformation}
\end{align}
The first equality is a definition, transforming an operator through a unitary is always through conjugation. The second equality is because, once the fermion is annihilated, there is nothing left to act on. Similarly,
\begin{align}
    G_{x+1}({b^n}^\dagger):= G_{x+1} {b^n}^\dagger G_{x+1}^\dagger = G_{x+1} {b^n}^\dagger = \sum_{n'} (g_{x+1})_{n'n}\, {b^{n'}}^\dagger. \label{eq:btransformation}
\end{align}
The demand that the fermionic transfer $T_R$ be $SU(2)$-gauge-covariant is the demand that
\begin{align}
    G(T_R)=T_R &\Leftrightarrow G T_R G^\dagger=T_R \label{eq:gaugesymtrans}\\
    &\Leftrightarrow G T_R =T_R G\\
    &\Leftrightarrow [G, T_R] =0 \label{eq:gaugesymcom}
\end{align}
The key point is that this must hold even when the $(G_x)_x$ are independent of each other. We immediately see that Eq.~\eqref{eq:TR} cannot be $SU(2)$-gauge-covariant, because

\begin{align}
    G(T_R)&=\sum_{m,n} {b^n}^\dagger\, g_{x+1}\,  M^{*}_{m n }\, g_x^\dagger\, a^m  \label{eq:transformTR}\\
    &\neq T_R \textrm{ in general}.
\end{align}

This is the way gauge theories are constructed. One starts by acknowledging the fact that the fermionic theory is not gauge-symmetric, and then tries to fix it. Let us give a rough outline of how to achieve this.

First of all, Eq.~\eqref{eq:TR} is kind of heavy and suggests that we introduce the notation

\begin{align}
    \bar{a} = (a^+\ a^-)^{T}\qquad\bar{b} = (b^+\ b^-)^T\quad\textrm{so that:}\quad\bar{b}^\dagger=({b^+}^\dagger {b^-}^\dagger) \label{eq:barconventions}
\end{align}

Then the equation rewrites as

\begin{align}
T_R = \bar{b}^\dagger\, \bar{M}^{\dagger}\, \bar{a} .\label{eq:TRbar}
\end{align}
Eqs.~\eqref{eq:atransformation} and \eqref{eq:btransformation} rewrite as

\begin{align}
G_x(\bar{a}) = \, g_x^\dagger\bar{a}, \qquad G_{x+1}(\bar{b}^\dagger) = \bar{b}^\dagger \, g_{x+1},\label{eq:barabtransformation}
\end{align}
where the products $\, g_x^\dagger\bar{a}$ and $\bar{b}^\dagger \, g_{x+1}$ are defined as
\begin{align}
(g_x^\dagger \bar{a})_{m'} &:= \sum_{m} a^{m}\, (g_x^\dagger)_{m' m} = \sum_{m} a^{m}\,(g_x)^*_{m m'},\\
( \bar{b}^\dagger g_{x+1}) _{n'} &:= \sum_{n} (g_{x+1})_{n n'}\,{b^n}^\dagger.\label{eq:fermtransf-convention}
\end{align}

It follows that Eq.~\eqref{eq:transformTR} rewrites as

\begin{align}
G(T_R) &=  \bar{b}^\dagger \, g_{x+1}\, \bar{M}^{\dagger} \, g_x^\dagger \, \bar{a} \label{eq:transformTRbar}\\
&\neq T_R.
\end{align}

Having introduced this notation, we see that it would be wonderful if we had some $\hat{M}^{\dagger}$ with the property that

\begin{align}
    G(\hat{M}^{\dagger}) = g_{x+1}^\dagger \, \hat{M}^{\dagger} g_x,  \label{eq:Mtransformation}
\end{align}
so that

\begin{align}
    G(T_R) &= \bar{b}^\dagger \, g_{x+1} \,  g_{x+1}^\dagger \, \hat{M}^{\dagger} g_x \, g_x^\dagger  \, \bar{a}  \label{eq:transformTRbarhat}\\
    &= \bar{b}^\dagger \, \hat{M}^{\dagger} \, \bar{a} \label{eq:slippery}\\
    &= T_R.
\end{align}
Step Eq.~\eqref{eq:slippery} is somewhat slippery and misleading, however, because: if the $M_{mn}^{*}$ are just numbers, the inner $g_x$ and $g_{x+1}^\dagger$ are not matrices, they cannot cancel out with the outer $g_x^\dagger$ and $g_{x+1}$; moreover if they did then $\bar{a}$, $\bar{b}^\dagger$ would no longer be transforming. On what system are the inner $g_x$ and $g_{x+1}^\dagger$ really acting? The solution will come from resolving this question and making this equation really formal. For this, we will have to elevate the numbers $M_{mn}^{*}$ into operators $\hat{M}_{mn}^{*}$ that act on extra degrees of freedom: the gauge field.

\section{The gauge field}\label{sec:gaugefield}

Let us quickly remind ourselves of how this works for $U(1)$ \cite{ArrighiQED}.
The group $U(1)$ is $\{e^{i\theta}\mid \theta\in\mathbb{R}\}$ endowed with multiplication. Recall that its representations are maps $\pi^l: U(1) \to GL(V_l)$ such that $\pi^l(g h) = \pi^l(g)\,\pi^l(h)$, with $V_l$ a finite-dimensional vector space and $GL(V_l)$ the corresponding group of invertible matrices. Clearly the maps $\pi^l:g\mapsto g^l$ are representations with $V_l\cong\mathbb{C}$. For instance $g=e^{i\theta}\in U(1)$ has the $l=2$ representation $\pi^2(g)=e^{i2\theta}\in GL(V_2)$, where $GL(V_2)$ stands for the invertible linear maps over $V_2$, i.e.\ elements of $V_2^*\otimes V_2$---incidentally in this case $V_2^*\otimes V_2\cong\mathbb{C}$ because $V_2\cong\mathbb{C}$.
These precise $\pi^l$ are the irreducible representations of $U(1)$ because all others are isomorphic to direct sums of these, by the Peter--Weyl theorem \cite{BrockerTomDieck1985,Sepanski2006-mh,hall2003lie}, cf.\ Appendix~\ref{app:peter-weyl}. By the same theorem, we learn how to use these irreducible representations to construct $L^2(U(1))$, namely

\begin{align}
    L^2(U(1))&\cong\bigoplus_{l\in\mathbb{Z}} V_l^*\otimes V_l\\
    &\cong \textrm{Span}\{\ket{l}\ket{l}\mid l\in\mathbb{Z}\}\label{eq:U1gaugefield}
\end{align}
Of course for $U(1)$ the use of this theorem is a bit of overkill. We knew already that the square-integrable wavefunctions $\psi: U(1)\to \mathbb{C}$ decompose in the Fourier basis as $\psi=\sum_l \tilde{\psi}(l) f^l$, with $f^l:=(g\mapsto g^l)$ and $\tilde{\psi}(l)=\int_{U(1)} dg\, (f^l(g))^* \psi(g)$. Therefore, the $l$ of Eq.~\eqref{eq:U1gaugefield} are nothing but the labels of the modes $f^l$ of $L^2(U(1))$.\\
For $U(1)$-gauge-covariant fermionic transfer \cite{ArrighiQED}, the link from $x$ to $x+1$ carries the gauge field, i.e.\ an instance of $L^2(U(1))$. More precisely, the source end of the link carries $\ket{l}^{x:R}$, and the target end carries $\ket{l}^{(x+1):L}$, matching the two tensor factors in Eq.~\eqref{eq:U1gaugefield}. The gauge field raising operator is defined by
\begin{align}
    V(\ket{l}\ket{l})  &= \ket{l+1}\ket{l+1} \label{eq:raiseU1gaugelabels}
\end{align}
We then let

\begin{align}
T^{U(1)}_R = b^\dagger V^{\dagger} a  .
\end{align}
It is very instructive to reframe the raising operator of Eq.~\eqref{eq:raiseU1gaugelabels} not in terms of increasing mode labels, but in terms of pointwise multiplication of modes. Indeed,
\begin{align}
    f^kf^l = (g\mapsto g^k)(g\mapsto g^l):=(g\mapsto g^kg^l)=(g\mapsto g^{k+l})=f^{k+l}\label{eq:multiplyU1gaugemodes}
\end{align}
For this reason, we may define
\begin{align}
&(\ket{k}\ket{k})\odot(\ket{l}\ket{l}):=\ket{k+l}\ket{k+l}\label{eq:multiplyU1gaugelabels}\\
\textrm{so that}\quad &(V\ \bullet)= ((\ket{1}\ket{1})\odot\ \bullet)\label{eq:multiplyU1gaugelabelsby1}
\end{align}
In order to construct an $SU(2)$-gauge-covariant fermionic transfer, we will proceed in the same manner. Each link will carry the gauge field, i.e.\ an instance of $L^2(SU(2))$.\\
For most quantum information scientists the group $SU(2)$ is the group of $2\times 2$ unitary matrices with determinant $1$, endowed with matrix multiplication, as generated by exponentials $e^{i\theta^a S_a}$, where $S_x=\sigma_x/2,\,S_y=\sigma_y/2,\,S_z=\sigma_z/2$ are the Hermitian spin operators built from the Pauli matrices $\sigma_x,\sigma_y,\sigma_z$, and obey the commutation relations $[S_a, S_b] = i\,\epsilon_{abc}\,S_c$. But in representation theory, $SU(2)$ is defined as the Lie group generated by the Lie algebra $[S_a, S_b] = i\,\epsilon_{abc}\,S_c$, whose representations yield, via exponentiation, the representations of the Lie group $SU(2)$. Following this approach \cite{hall2003lie}, one finds that for each half-integer $j\in\mathbb{N}/2$, there is an irreducible representation $\pi^j$ with $V_j\cong\mathbb{C}^{2j+1}$; these are all the irreducible representations of $SU(2)$, all others being isomorphic to direct sums of these, by the Peter--Weyl theorem. Again by the same theorem, we learn that

\begin{align}
    L^2(SU(2))&\cong\bigoplus_{j=0,\tfrac12,1,\ldots} V_j^*\otimes V_j\label{eq:SU2gaugefield}\\
    &\cong \textrm{Span}\{\ket{jm}\ket{jn}\mid j=0,\tfrac12,1,\ldots,\ m,n\in\{-j\ldots j\}\}
\end{align}
Now the modes of $L^2(SU(2))$ are the Wigner $D$-matrix elements $D^j_{mn}:=(g\mapsto \bra{jm}\pi^j(g)\ket{jn})$. Indeed, by the Peter--Weyl theorem, these are orthogonal:
\begin{align}
    \int_{SU(2)} dg\, (D^{j'}_{m'n'}(g))^* D^j_{mn}(g) = \frac{\delta_{jj'}\delta_{mm'}\delta_{nn'}}{2j+1}\label{eq:Dorthogonality}
\end{align}
and generate the whole space: the $\sqrt{2j+1}\,D^j_{mn}$ form an orthonormal basis of $L^2(SU(2))$.
Thus any $\psi\in L^2(SU(2))$ decomposes as $\psi=\sum_{j,m,n} \tilde{\psi}(j,m,n)\, \sqrt{2j+1}\,D^j_{mn}$, with $\tilde{\psi}(j,m,n)=\sqrt{2j+1}\int_{SU(2)} dg\, (D^j_{mn}(g))^*\, \psi(g)$. Therefore, the kets $\ket{jm}\ket{jn}$ of Eq.~\eqref{eq:SU2gaugefield} are the labels of the orthonormal modes $\sqrt{2j+1}\,D^j_{mn}$, and the $\tilde{\psi}(j,m,n)$ are their amplitudes.
Recall that for $U(1)$, the gauge field raising operator was reframed as stemming from the pointwise multiplication of modes, see Eq.~\eqref{eq:multiplyU1gaugelabelsby1}. Similarly, the pointwise product of $SU(2)$ modes is \cite{VMK}
\begin{align}
    D^j_{mn}D^{j'}_{m'n'}=(g\mapsto D^j_{mn}(g))(g\mapsto D^{j'}_{m'n'}(g)) &:= (g\mapsto D^j_{mn}(g)\, D^{j'}_{m'n'}(g)).\label{eq:multiplySU2gaugemodes}
\end{align}
Because the RHS is itself an element of $L^2(SU(2))$, it can again be expressed as a sum of $D^j_{mn}$ modes. We recall it here just for completeness, but our later unitarity and gauge covariance proofs will not need it:
\begin{align}
    D^j_{mn}D^{j'}_{m'n'} = \sum_{k=|j-j'|}^{j+j'} \braket{jm, j'm'}{k\, (m\!+\!m')} \braket{jn, j'n'}{k\, (n\!+\!n')}\, D^k_{m+m',n+n'}\label{eq:multiplySU2gaugemodeswithCG}
\end{align}
where $\braket{jm, j'm'}{k\,q}$ are Clebsch--Gordan coefficients, see Eq.~\eqref{eq:multiplySU2gaugemodeswithCG} of Appendix~\ref{app:CG}. Re-expressing in terms of the orthonormal mode labels, we have
\begin{align}
    &\ket{jm}\ket{jn}\odot\ket{j'm'}\ket{j'n'}\\
    =&\sum_{k=|j-j'|}^{j+j'} \sqrt{\frac{(2j\!+\!1)(2j'\!+\!1)}{2k\!+\!1}}\, \braket{jm, j'm'}{k\, (m\!+\!m')} \braket{jn, j'n'}{k\, (n\!+\!n')}\, \ket{k\,(m\!+\!m')}\ket{k\,(n\!+\!n')}\label{eq:multiplySU2gaugelabels}
\end{align}
In analogy with Eq.~\eqref{eq:multiplyU1gaugelabelsby1}, define the $SU(2)$ comparator entry $\hat{M}_{mn}$ as multiplication by the fundamental mode:
\begin{align}
(\hat{M}_{mn}\ \bullet) := \tfrac{1}{\sqrt{2}}\,(\ket{\frac{1}{2}m}\ket{\frac{1}{2}n}\odot\ \bullet)\label{eq:multiplySU2gaugelabelsbymn}
\end{align}
The prefactor $\tfrac{1}{\sqrt{2}}=\tfrac{1}{\sqrt{2j+1}}$ at $j=\tfrac12$ undoes the normalization of the mode label, so that $\hat{M}_{mn}$ is pointwise multiplication by the fundamental mode $D^{1/2}_{mn}$ itself. Then $\hat{M}$ really is multiplication by the matrix $g$ at which the wavefunction is evaluated, in the sense that $\pi^{1/2}(g)=g$ and hence $D^{1/2}_{mn}(g)=g_{mn}$ exactly, so that on a wavefunction $\psi_\pm\in L^2(SU(2))$ carrying a colour index,
\begin{align}
    \hat{M}\psi:=\bigl(g\mapsto g\,\psi(g)\bigr)\label{eq:Mismultbyg}
\end{align}
i.e. a $2\times 2$ matrix product on the colour index, pointwise in $g$. Since $g$ is unitary, $\hat{M}$ preserves the norm: it is everywhere defined on $L^2(SU(2))$, no matter that $\odot$ itself is not.

\section{The gauge transformation}\label{sec:gaugetransformations}

We now aim to describe how each $G_x$ acts.
Remember that each site $x$ holds fermions, but it also has an $x:L$ half-link on its left, and an $x:R$ half-link on its right. Roughly speaking, the gauge transformation works by applying, to each of them, the correct $j$ representation of $g\in SU(2)$.
\begin{align}
G_x &:=\gamma^{x:L}\otimes\left(\bigotimes_{\mathrm{each\ fermion}}\gamma\right)\otimes \gamma^{x:R}\label{eq:Gxdef}\\
\textrm{with }\quad \gamma &:= \bigoplus_j \pi^j(g)
\end{align}
with two caveats: (i) the $\gamma$ on fermionic degrees of freedom can be truncated to $j=0, 1/2$ and must be expressed in the occupation number basis, and (ii) the $\gamma$ on the link labels transform as if acting left and right of matrix entries, see below.

First let us quickly remind ourselves of how this works for $U(1)$.
Starting with fermions, because the first non-trivial representation of the $U(1)$ group lives in $V_1^*\otimes V_1$, each fermion carries, besides its chirality, a $1$-dimensional degree of freedom upon which to act with $\pi^1(g)=g$. In other words,
\begin{align*}
\gamma\ket{k'k}^x = g^{k'+k}\ket{k'k}^x
\end{align*}
implying that Eqs.~\eqref{eq:atransformation} and \eqref{eq:btransformation} are met, namely
\begin{align*}
G_x(a_x) = a_x\ g_x^{-1},\qquad G_x(b_x^\dagger) = g_x\ b_x^\dagger.
\end{align*}
Indeed, on the non-zero cases,

\begin{align*}
G_x a_x G_x^\dagger\ket{1k}^x
&=g_x^{-(k+1)}G_x\ket{0k}^x
=g_x^{-(k+1)}g_x^{k}\ket{0 k}^x
=g_x^{-1}a_x\ket{1 k}^x,\\
G_x b_x^\dagger G_x^\dagger\ket{k' 0 }^x
&=g_x^{-k'}G_x\ket{k' 1}^x
=g_x^{-k'}g_x^{1+k'}\ket{ k' 1 }^x
=g_xb_x^\dagger\ket{k' 0 }^x.
\end{align*}

For the right half-link, associated to $V_l^{*}$:
\begin{align*}
\gamma^{x:R}\ket{l}^{x:R} &:= \gamma^*\ket{l}^{x:R} = g^{-l}\ket{l}^{x:R}
\end{align*}
For the left half-link, associated to $V_l$:
\begin{align*}
\gamma^{x:L}\ket{l'}^{x:L} &:= \gamma\ket{l'}^{x:L} = g^{l'}\ket{l'}^{x:L}.
\end{align*}
All this implies that $V$ transforms as Eq.~\eqref{eq:Mtransformation}, namely

\begin{align*}
G(V) = g_x^{-1}\ V\ g_{x+1}.
\end{align*}
Indeed,
\begin{align*}
G V G^\dagger\ket{l}^{x:R}\ket{l}^{(x+1):L}=g_x^{-(l+1)}g_{x+1}^{(l+1)}g_x^{l}g_{x+1}^{-l}\ket{l+1}^{x:R}\ket{l+1}^{(x+1):L}=g_x^{-1} g_{x+1}\,V\ket{l}^{x:R}\ket{l}^{(x+1):L}.
\end{align*}
Overall $G_x$ is defined by linear extension of
\begin{align*}
G_x\ket{l'}^{x:L}\ket{k'k}^x\ket{l}^{x:R}:=g^{l'+k'+k-l}\ket{l'}^{x:L}\ket{k'k}^x\ket{l}^{x:R}.
\end{align*}

Let us now do the same for $SU(2)$.
Because the first non-trivial representation of the $SU(2)$ group is the fundamental representation $V_{1/2}$, each fermion carries, besides its chirality, a $2$-dimensional `colour' degree of freedom upon which $\gamma$ acts with $\pi^{1/2}(g)=g$.
More explicitly, each site $x$ now holds four qubits, corresponding to the occupation-number representation of $b_x^+, b_x^-, a_x^+, a_x^-$:
\begin{align*}
\ket{k_{l}^+\,k_{l}^-\,k_{r}^+\,k_{r}^-}^x =: \ket{\bold{k}_{l} \bold{k}_{r}}^x, \quad\text{ with } k_{r}^\pm,\,k_{l}^\pm \in \{0,1\} \text{ and } \bold{k}_{r},\bold{k}_{l}\in\{0,1\}^2.
\end{align*}
The $b$ slots are placed to the left of the $a$ slots so that the pair acted upon by the on-site swap $S$, and the pair acted upon by the cross-site transfer $T$ (Sec.~\ref{sec:unitarity}), are both pairs of neighbouring slots.
We introduce the notations
\begin{align}
    \ket{+}:=\ket{1,0} \qquad \ket{-}:=\ket{0,1} \label{eq:plusminusstates}
\end{align}
On each of $\ket{\bold{k}_{l}}$ $\ket{\bold{k}_{r}}$, $\gamma$ acts as
\begin{align}
\gamma\ket{0,0} &:= \ket{0,0} \label{eq:gammafermions}\\
\gamma\ket{1,1} &:= \ket{1,1}\\
\gamma\ket{n} &= \sum_m (g_x)_{mn}\,\ket{m} \quad\text{for $m,n\in\{+,-\}$} \label{eq:gammasingle}.
\end{align}
As for the links, we must think of the $\ket{jm}^{x:R}\ket{jn}^{(x+1):L}$ as labelling some link matrix entries $L_{mn}$ (not as some $\ket{jm}\bra{jn}$). A gauge transformation at the source should multiply this link matrix $L$ on the left but daggered, and a gauge transformation at the target should multiply it on the right:
\begin{align*}
L\mapsto g_x^\dagger\,L\, g_{x+1}.
\end{align*}
Hence
\begin{align*}
(\pi^j(g)^\dagger L)_{mn}&=\sum_{m'}\pi^j(g)^\dagger_{mm'}L_{m'n}=\sum_{m'}\pi^j(g)^*_{m'm}L_{m'n},&
(L\pi^j(g))_{mn'}&=\sum_{n}L_{mn}\pi^j(g)_{nn'}.
\end{align*}
Consequently, the row label at $x:R$ and the column label at $x:L$ transform as
\begin{align}
    \gamma^{x:R}\ket{jm}^{x:R} &= \sum_{m'}\pi^j(g)^\dagger_{mm'}\,\ket{jm'}^{x:R} = \sum_{m'}\pi^j(g)^*_{m'm}\,\ket{jm'}^{x:R}\label{eq:halflinkR}\\
    \gamma^{x:L}\ket{j'n'}^{x:L} &= \sum_{n}\pi^{j'}(g)_{nn'}\,\ket{j'n}^{x:L}\label{eq:halflinkL}
\end{align}
Overall $G_x$ is defined by linear extension of

\begin{align}
G_x\ket{j'n'}^{x:L}\ket{\bold{k}_{l} \bold{k}_{r}}^x\ket{jm}^{x:R} := \gamma^{x:L}\ket{j'n'}^{x:L}\otimes \\big(\gamma\ket{\bold{k}_{l} }\,\gamma\ket{\bold{k}_{r}}\big)^x\otimes \gamma^{x:R}\ket{jm}^{x:R}.\label{eq:overallGx}
\end{align}
Notice that \cite{VMK}:
\begin{equation}
\pi^j(g)^*_{mm'}= (-1)^{m-m'}\,\pi^j(g)_{-m,-m'}.\label{eq:SU2pistar}
\end{equation}

\section{Gauge-covariant transfer}\label{sec:covariance}

We now prove that, with $\hat M$ as in Eq.~\eqref{eq:multiplySU2gaugelabelsbymn} and $T_R=\bar b_{x+1}^\dagger\,\hat M^\dagger\,\bar a_x$ (Eq.~\eqref{eq:TRbar}), the gauge covariance $G(T_R)=T_R$ of Eq.~\eqref{eq:slippery} holds. Throughout, $g_x\in SU(2)$ and $G_x$ are defined from one another as in Eq.~\eqref{eq:overallGx} and $G$ as in Eq.~\eqref{eq:G}. Within $G$, only $G_x,G_{x+1}$ touch the support of $T_R$. On the link, they act as $G_{x,x+1}:=\gamma^{x:R}\otimes\gamma^{(x+1):L}$. Hence
\begin{align}
G(T_R)\;=\;G_{x+1}(\bar b_{x+1}^\dagger)\,G_{x,x+1}(\hat M^\dagger)\,G_x(\bar a_x).\label{eq:GTRfactor}
\end{align}

\noindent {\em Fermionic gauge transformation.}
We first prove Eq.~\eqref{eq:barabtransformation}, namely
\begin{align}
G_x(\bar a_x)\;=\;g_x^\dagger\,\bar a_x,\qquad G_{x+1}(\bar b_{x+1}^\dagger)\;=\;\bar b_{x+1}^\dagger\,g_{x+1}.\label{eq:fermtransf}
\end{align}
\begin{proof} $G_x(\bar a_x)$ reduces to the conjugation of $\bar a_x$ by the $\gamma$ acting on $\ket{\bold{k}_{r}}$ as in Eq.~\eqref{eq:gammafermions}. Hence we must check that
\begin{align}
\gamma\,a^m\,\gamma^\dagger\;=\;\sum_{m'}(g_x)^*_{m'm}\,a^{m'}\label{eq:gammaa}
\end{align}
holds on each basis state:
\begin{itemize}
\item $\gamma\,a^m\,\gamma^\dagger\,\ket{0,0}=0=\sum_{m'}(g_x)^*_{m'm}\,a^{m'}\,\ket{0,0}$.
\item On $\ket{m'}$ for $m'\in\{+,-\}$:
\begin{align*}
\gamma\,a^m\,\gamma^\dagger\,\ket{m'}\;=\;\gamma\,a^m\,{\sum_{m''}}(g_x)^*_{m'm''}\ket{m''}\;=\;(g_x)^*_{m'm}\,\ket{0,0}\; =\; {\sum_{m''}}(g_x)^*_{m''m}\,a^{m''}\,\ket{m'}.
\end{align*}
\item Due to anticommutation, $a^\pm\ket{1,1}=\pm\ket{\mp}$. Then
\begin{align*}
\textrm{LHS}\;&=\;\gamma\,a^m\,\gamma^\dagger\,\ket{1,1}\;=\;\pm\,\gamma\,\ket{\mp}\;=\;\pm(g_x)_{+\mp}\,\ket{+}\pm(g_x)_{-\mp}\,\ket{-}\\
\textrm{RHS}\;&=\;(g_x)^*_{+\pm}\,\ket{-}-(g_x)^*_{-\pm}\,\ket{+}.
\end{align*}
Equating coefficients we need 1/ for $\ket{+}$, $\pm(g_x)_{+\mp}=-(g_x)^*_{-\pm}$, and 2/ for $\ket{-}$, $\pm(g_x)_{-\mp}=(g_x)^*_{+\pm}$. Both follow from $(g_x)_{--}=(g_x)^*_{++}$ and $(g_x)_{+-}=-(g_x)^*_{-+}$, which in turn come from $g_x\in SU(2)$.
\end{itemize}
The same case check at site $x+1$ on $b^n$ gives $\gamma\,b^n\,\gamma^\dagger=\sum_{n'}(g_{x+1})^*_{n'n}\,b^{n'}$. Hence $\gamma\,{b^n}^\dagger\,\gamma^\dagger=\sum_{n'}(g_{x+1})_{n'n}\,{b^{n'}}^\dagger$, which is the second equality.
\end{proof}
\noindent {\em Link gauge transformation.}
Second, we prove Eq.~\eqref{eq:Mtransformation}, namely
\begin{align}
G_{x,x+1}(\hat M)\;=\;g_x^\dagger\,\hat M\,g_{x+1}.\label{eq:comptransf}
\end{align}
\begin{proof}
\begin{align}
G_{x,x+1}\,\ket{jm}^{x:R}\ket{jn}^{(x+1):L}
\;&=\;\sum_{m'n'}\pi^j(g_x)^*_{m'm}\,\pi^j(g_{x+1})_{n'n}\,\ket{jm'}^{x:R}\ket{jn'}^{(x+1):L}&&\text{(Eqs.~\eqref{eq:halflinkR},\eqref{eq:halflinkL})}\nonumber\\
G_{x,x+1} (g\mapsto \pi^j(g)_{mn}) &= g\mapsto\sum_{m'n'}\pi^j(g_x)^*_{m'm}\,\pi^j(g_{x+1})_{n'n}\,\bra{jm'}\pi^j(g)\ket{jn'}&&\text{(as modes)}\nonumber\\
&=g\mapsto\bra{jm}\pi^j(g_x^{-1})\,\pi^j(g)\,\pi^j(g_{x+1})\ket{jn}&&\text{(by unitarity)}\nonumber\\
&=g\mapsto\pi^j\!\bigl(g_x^\dagger\,g\,g_{x+1}\bigr)_{mn}.\label{eq:bitranslation}
\end{align}
where the last line is because $\pi^j$ is a representation.
Now consider
\begin{align}
    G_{x,x+1}\bigl(\ket{jm}\ket{jn}\odot\ket{j'm'}\ket{j'n'}\bigr)\label{eq:transformodot}
\end{align}
Recall the $\odot$ product on mode labels was defined by switching to modes, performing the pointwise product, and switching back to labels (Eqs.~\eqref{eq:multiplySU2gaugemodes},\eqref{eq:multiplySU2gaugelabels}). In terms of modes, Eq.~\eqref{eq:transformodot} reads
\begin{align*}
G_{x,x+1}\bigl(g\mapsto \pi^j(g)_{mn}\,\pi^{j'}(g)_{m'n'}\bigr)
&= g\mapsto \pi^j(g_x^\dagger g g_{x+1})_{mn}\,\pi^{j'}(g_x^\dagger g g_{x+1})_{m'n'}&&\text{(Eq.~\eqref{eq:bitranslation})}
\end{align*}
Switching back to labels this is
\begin{align*}
\bigl(G_{x,x+1}\ket{jm}\ket{jn}\bigr)\odot\bigl(G_{x,x+1}\ket{j'm'}\ket{j'n'}\bigr)
\end{align*}
Hence,
\begin{align}
G_{x,x+1}(\psi\odot\phi)\;=\;(G_{x,x+1}\,\psi)\odot(G_{x,x+1}\,\phi)\label{eq:odotcov}
\end{align}
for $\psi,\phi$ finite combinations of modes, which is all we shall need.


Apply this to $\hat M_{mn}\,\bullet:=\tfrac{1}{\sqrt2}\,\ket{\tfrac12 m}^{x:R}\ket{\tfrac12 n}^{(x+1):L}\odot\,\bullet$ (Eq.~\eqref{eq:multiplySU2gaugelabelsbymn}):
\begin{align*}
G_{x,x+1}(\hat M_{mn})\,\bullet
&=G_{x,x+1}\,\hat M_{mn}\,G_{x,x+1}^\dagger\,\bullet&&\\
&=\tfrac{1}{\sqrt2}\,G_{x,x+1}\bigl(\ket{\tfrac12 m}^{x:R}\ket{\tfrac12 n}^{(x+1):L}\odot G_{x,x+1}^\dagger\bullet\bigr)&&\text{(Eq.~\eqref{eq:multiplySU2gaugelabelsbymn})}\\
&=\tfrac{1}{\sqrt2}\,\bigl(G_{x,x+1}\ket{\tfrac12 m}^{x:R}\ket{\tfrac12 n}^{(x+1):L}\bigr)\odot(G_{x,x+1} G_{x,x+1}^\dagger\bullet)&&\text{(Eq.~\eqref{eq:odotcov})}\\
&=\tfrac{1}{\sqrt2}\,\bigl(G_{x,x+1}\ket{\tfrac12 m}^{x:R}\ket{\tfrac12 n}^{(x+1):L}\bigr)\odot \bullet&&\text{(unitarity)}\\
&=\tfrac{1}{\sqrt2}\,\bigl(\sum_{m'n'}(g_x^\dagger)_{mm'}(g_{x+1})_{n'n}\ket{\tfrac12 m'}^{x:R}\ket{\tfrac12 n'}^{(x+1):L}\bigr)\odot \bullet &&\text{(Eqs.~\eqref{eq:halflinkR},\eqref{eq:halflinkL})}\\
&=\sum_{m'n'}(g_x^\dagger)_{mm'}(g_{x+1})_{n'n}\hat M_{m'n'}\bullet &&\text{(Eq.~\eqref{eq:multiplySU2gaugelabelsbymn}).}
\end{align*}
Reading off the $(m,n)$ entry,
\begin{align*}
G_{x,x+1}(\hat M)_{mn}\;=\;\sum_{m'n'}(g_x^\dagger)_{mm'}\,\hat M_{m'n'}\,(g_{x+1})_{n'n}\;=\;(g_x^\dagger\,\hat M\,g_{x+1})_{mn}.
\end{align*}
\end{proof}

\noindent {\em Overall.}
\begin{align*}
G(T_R)
&=\bar b_{x+1}^\dagger\,g_{x+1}\,\cdot\,g_{x+1}^\dagger\,\hat M^\dagger\,g_x\,\cdot\,g_x^\dagger\,\bar a_x&&\text{by \eqref{eq:fermtransf},\eqref{eq:comptransf}}\\
&=\bar b_{x+1}^\dagger\,(g_{x+1}g_{x+1}^\dagger)\,\hat M^\dagger\,(g_xg_x^\dagger)\,\bar a_x&&\\
&=\bar b_{x+1}^\dagger\,\hat M^\dagger\,\bar a_x&&\\
&=T_R.&&
\end{align*}
Each cancellation $g_{x+1}g_{x+1}^\dagger=\mathbb{I}_2$ (resp. $g_xg_x^\dagger=\mathbb{I}_2$) is between $2\times 2$ matrices in $SU(2)$, corresponding to the colour indices of $\bar a_x,\bar b_{x+1}^\dagger$ and $\hat M$. This makes Eq.~\eqref{eq:slippery} fully rigorous: the inner $g_{x+1}^\dagger,g_x$ are extracted by Eq.~\eqref{eq:comptransf} on $\hat M^\dagger$ (the dagger of Eq.~\eqref{eq:comptransf}), the outer $g_{x+1},g_x^\dagger$ by Eq.~\eqref{eq:fermtransf} on $\bar b_{x+1}^\dagger,\bar a_x$.

\section{$SU(2)$-gauged Dirac QW and its unitarity}\label{sec:unitarity}

This section presents the one-particle sector of the $SU(2)$-gauged Dirac QCA, i.e.\ the $SU(2)$-gauged Dirac quantum walk (QW). The multi-particle extension is deferred to Sec.~\ref{sec:multi-particle}. In the one-particle sector, each qubit encodes the occupation of a fermionic mode, and exactly one qubit is in $\ket{1}$.

\noindent{\em Dirac QW.}
The Dirac QW \cite{ArrighiNesmeForets2014,ArrighiQED} is a discrete-space, discrete-time unitary scheme that simulates the Dirac equation on a quantum computer. Fermions propagate through three sub-steps per timestep: (1)~an on-site swap $S$ exchanging chiralities $a\leftrightarrow b$; (2)~a cross-site transport $T$ moving the fermion across the link to the neighbouring site; (3)~an on-site mass rotation $C$ mixing chiralities by an angle $\epsilon m$. One timestep is $W=C\cdot T\cdot S$. Its convergence to the Dirac equation is by now well established \cite{ArrighiNesmeForets2014}.

Now that $S$ is on the table we can settle who moves where. Overall, over a whole timestep, it is the $b$ fermions that move right. Indeed, they start in the $b$ slots, on the left of the site; $S$ swaps them for the $a$ fermions, i.e.\ into the $a$ slots, which are the ones the transfer treats as right-movers; and $T$ then carries them into the $b$ slots of the site $x+1$. So they do move right, and they are back in a $b$ slot, ready for the next timestep. Symmetrically, the $a$ fermions move left, since $S$ puts them in the $b$ slots and $T$ carries those into the $a$ slots of the site $x-1$. In short, `right-mover' names the $a$ slots at the transfer sub-step (Eq.~\eqref{eq:TRbar}), and the $b$ slots over a full timestep. Both $S$ and $T$ being swaps, only their composition $T\cdot S$ is a displacement that preserves chirality.

\noindent{\em $U(1)$-gauged Dirac QW.}
For $U(1)$-gauge covariance \cite{ArrighiQED}, each link carries $L^2(U(1))$ (Eq.~\eqref{eq:U1gaugefield}), and the cross-site transport is modified to update the gauge field via $V$ (Eq.~\eqref{eq:raiseU1gaugelabels}). Each site $x$ holds two qubits for $b_x,a_x$. In the canonical basis $\{\ket{00},\ket{01},\ket{10}\}^x$ (first qubit~$b_x$, second~$a_x$), the on-site operators are
\begin{align}
S_x = 1\oplus\begin{pmatrix} 0&1\\1&0\end{pmatrix},\qquad
C_x = 1\oplus\begin{pmatrix} c&-is\\-is&c\end{pmatrix},\label{eq:SCu1}
\end{align}
with $c=\cos(\epsilon m),\,s=\sin(\epsilon m)$. The swap $S_x$ exchanges $\ket{01}\leftrightarrow\ket{10}$. The mass coin $C_x$ rotates chirality within the one-particle sector. The cross-site transport $T_x$ acts on the pair $(b_{x+1},a_x)$ and the link $x\to x+1$. In the one-particle sector:
\begin{align}
T_x = 1\oplus\begin{pmatrix} 0&V_x^\dagger\\V_x&0\end{pmatrix},\label{eq:Tu1}
\end{align}
where $V_x,V_x^\dagger$ are operators on $L^2(U(1))$: right-mover $a_x$ hops to $b_{x+1}$ while raising the gauge field by $V_x$, and vice versa with $V_x^\dagger$.

Each sub-step is built from $a,b,V$, which transform under $G$ as Eqs.~\eqref{eq:atransformation},\eqref{eq:btransformation},\eqref{eq:Mtransformation} (with $V$ in place of $\hat M$). Gauge covariance of each sub-step follows by the same cancellations as in Sec.~\ref{sec:covariance}. Unitarity of $W=C\cdot T\cdot S$ follows from composing unitaries: $S$ and $C$ are manifestly unitary, and $T$ is unitary because $V$ is (it is a shift on $L^2(U(1))$).

\noindent{\em $SU(2)$-gauged Dirac QW.}
For $SU(2)$, each site holds four qubits $b^+,b^-,a^+,a^-$ (Sec.~\ref{sec:gaugetransformations}) and each link carries $L^2(SU(2))$ (Eq.~\eqref{eq:SU2gaugefield}). In the canonical basis $\{\ket{0000},\ket{0001},\ket{0010},\ket{0100},\ket{1000}\}^x$ of the $0$- and $1$-particle sectors of $\ket{k_{l}^+\,k_{l}^-\,k_{r}^+\,k_{r}^-}^x$, the three sub-steps are the minimal colour extension of the $U(1)$ case. Here and throughout, basis states are ordered by standard binary counting of the qubit string, with the rightmost qubit as the least significant bit. In particular, when a one-particle block is said to act on a pair of modes, the pair is listed in that same order, the least significant mode of the two coming first.

$S_x$ is the on-site chirality swap, tensored with the identity on colour:
\begin{align}
S_x = (1\oplus\ S^{1}_{x}), \hspace{4mm} S^{1}_{x} = S \otimes \mathbb{I}_2, \hspace{4mm} S = \begin{pmatrix} 0&1\\1&0\end{pmatrix}.\label{eq:SSU2}
\end{align}
$S_x$ commutes with $G$ (Eq.~\eqref{eq:overallGx}) because it acts as the identity on colour and on the link, hence is gauge-covariant.

$C_x$ is the mass rotation, tensored with the identity on colour:
\begin{align}
C_x = (1\oplus C_{x}^{1} ), \hspace{4mm}  C_{x}^{1} = C \otimes \mathbb{I}_2, \hspace{4mm} C =  \begin{pmatrix} c& -is\\-is&c\end{pmatrix}.\label{eq:CSU2}
\end{align}
It is gauge-covariant by the same argument.

$T_x$ is cross-site transport with the comparator $\hat M$ (Eq.~\eqref{eq:multiplySU2gaugelabelsbymn}) in place of $V$. In the one-particle sector, on $(\bar b_{x+1},\bar a_x)$:

\begin{align}
T_x = (1\oplus T_{x}^{1}), \hspace{4mm} T_{x}^{1} = \begin{pmatrix} 0 & \hat M^\dagger \\ \hat M & 0 \end{pmatrix},\label{eq:TSU2}
\end{align}
where the upper-right block $\hat M^\dagger$ comes from $T_R=\bar b_{x+1}^\dagger\,\hat M^\dagger\,\bar a_x$ (Eq.~\eqref{eq:TRbar}), mapping $\ket{m}_a\mapsto\sum_n(\hat M^\dagger)_{nm}\ket{n}_b$, and the lower-left block $\hat M$ from $T_R^\dagger=\bar a_x^\dagger\,\hat M\,\bar b_{x+1}$, mapping $\ket{n}_b\mapsto\sum_m \hat M_{mn}\ket{m}_a$. This is Eq.~\eqref{eq:Tu1} with $\hat M$ in place of $V$. Since $T_x$ is built from $\bar a,\hat M,\bar b$, gauge covariance follows from Sec.~\ref{sec:covariance}.

Unitarity of $W=C\cdot T\cdot S$ reduces to unitarity of $T$, since that of $S$ and $C$ is manifest. From Eq.~\eqref{eq:TSU2}:
\begin{align*}
T_x^\dagger\, T_x \;=\;    1\oplus\begin{pmatrix} 0 & \hat M^\dagger \\ \hat M & 0 \end{pmatrix}^{\!2} \;=\; 1\oplus\begin{pmatrix} \hat M^\dagger\hat M & 0 \\ 0 & \hat M\hat M^\dagger \end{pmatrix}.
\end{align*}
Hence $T_x$ is unitary iff both $\hat M^\dagger\hat M=\mathbb{I}_2$ and $\hat M\hat M^\dagger=\mathbb{I}_2$, i.e.\
\begin{align}
(\hat M^\dagger\hat M)_{mm'}\;=\;\sum_n (\hat M_{nm})^\dagger\,\hat M_{nm'}\;=\;\delta_{mm'}\,\mathbb{I}_2.\label{eq:MMdagger}
\end{align}
\begin{proof}
By the conjugation identity at the end of Sec.~\ref{sec:gaugetransformations}, the operator adjoint of $\hat M_{mn}$ (multiplication by $g\mapsto g_{mn}$) is $(\hat M_{mn})^\dagger=(-1)^{m-n}\hat M_{-m,-n}$. Substituting into Eq.~\eqref{eq:MMdagger}:
$$\sum_n(-1)^{m-n}\hat M_{-n,-m}\,\hat M_{nm'}.$$

\noindent{\em Off-diagonal ($m\neq m'$):} each term is a commutator of multiplication operators on $L^2(SU(2))$, which vanishes because pointwise products of functions commute. For instance $(m,m')=(+,-)$ gives
\begin{align*}
\hat M_{--}\,\hat M_{+-} \;-\; \hat M_{+-}\,\hat M_{--} \;=\; 0.
\end{align*}

\noindent{\em Diagonal ($m=m'$):} both cases $(+,+)$ and $(-,-)$ yield
\begin{align}
\hat M_{++}\,\hat M_{--} \;-\; \hat M_{+-}\,\hat M_{-+}.\label{eq:detM}
\end{align}
Let us show that Eq.~\eqref{eq:detM} is the identity by having it act on an arbitrary $\ket{j'm'}\ket{j'n'}$.
Recall that $\hat M_{mn}\ket{j'm'}\ket{j'n'}=\tfrac{1}{\sqrt{2}}\,\ket{\tfrac{1}{2}m}\ket{\tfrac{1}{2}n}\odot\ket{j'm'}\ket{j'n'}$, which is defined by switching to modes and doing pointwise multiplication (Eq.~\eqref{eq:multiplySU2gaugelabelsbymn}). Writing $\sqrt{2j'\!+\!1}\,\pi^{j'}(g)_{m'n'}$ for the mode labelled by $\ket{j'm'}\ket{j'n'}$, and expressing Eq.~\eqref{eq:detM} as acting on modes, we have:
\begin{align*}
g\mapsto(g_{++}g_{--}-g_{+-}g_{-+})\sqrt{2j'\!+\!1}\,\pi^{j'}(g)_{m'n'}=g\mapsto\det(g)\sqrt{2j'\!+\!1}\,\pi^{j'}(g)_{m'n'}=g\mapsto\sqrt{2j'\!+\!1}\,\pi^{j'}(g)_{m'n'},
\end{align*}
where the last step is because $g\in SU(2)$ is \emph{special} unitary: $\det(g)=1$ by definition. Switching back to labels, this is $\ket{j'm'}\ket{j'n'}$, so Eq.~\eqref{eq:detM} is the identity.

\noindent{\em The other identity:} $(\hat M\hat M^\dagger)_{mm'}=\sum_n \hat M_{mn}\,(\hat M_{m'n})^\dagger$ differs from the above only by the order of the two factors in each term. Since the $\hat M_{mn}$ are multiplication operators they all commute, so this is the very same computation, and $\hat M\hat M^\dagger=\mathbb{I}_2$ as well. Note that this second identity does need to be checked: the $\hat M_{mn}$ act on the infinite-dimensional $L^2(SU(2))$, where $\hat M^\dagger\hat M=\mathbb{I}_2$ alone would not suffice.
\end{proof}

\section{Multi-particle case}\label{sec:multi-particle}
In this section, we construct the multi-particle fermionic evolution of the $SU(2)$-gauge-covariant Dirac quantum cellular automaton (QCA) by building upon the single-particle $SU(2)$-gauge-covariant Dirac quantum walk (QW) defined in Sec.~\ref{sec:unitarity}. This non-interacting multi-particle extension is valid because we do not consider explicit $N$-body fermionic interactions; instead, the fermions only couple minimally to the background gauge fields \cite{montvay1994quantum} and through their anticommutation.

Before proceeding, we establish several general mathematical results concerning non-interacting multi-particle extensions of fermionic evolutions. While these foundational concepts can be found scattered across various texts on many-body physics and matrix analysis \cite{fetter2003quantum, horn2012matrix}, here we present them in a synthetic, self-contained and rigorous manner.

Consider an $N$-mode fermionic system generated by the creation operators $\{a_1^\dagger, \dots, a_N^\dagger\}$, satisfying the standard fermionic anticommutation relations $\{a_i^\dagger, a_j^\dagger\} = 0$ and $\{a_i, a_j^\dagger\} = \delta_{ij}$.
We represent a $K$-particle basis state by an {\em index subset} $I \subseteq \{1, \dots, N\}$ of size $|I| = K$, understood as a list $(i_k)_k$ where $i_1 < i_2 < \dots < i_K$. Or equivalently, by its {\em occupation number representation state} $\mathbf{l} \in \{0, 1\}^N$, where $l_i = 1$ if $i \in I$ and $l_i = 0$ otherwise. The state is constructed by acting on the vacuum $\ket{\text{vac}}$ with the creation operators in strictly ascending order:
\begin{equation}
    \ket{\mathbf{l}} = (a_1^\dagger)^{l_1} (a_2^\dagger)^{l_2} \dots (a_N^\dagger)^{l_N} \ket{\text{vac}},
\end{equation}
where $(a_i^\dagger)^0 \equiv \mathbb{I}$, ensuring that the $i$-th creation operator is applied if and only if $i \in I$.

Let $\mathbf{U}$ be a unitary operator representing the multi-particle extension of a single-particle unitary evolution matrix $U$. We assume the vacuum state is invariant under this evolution ($\mathbf{U}\ket{\text{vac}} = \ket{\text{vac}}$). The fact that $\mathbf{U}$ is assumed to be non-interacting is enforced by the following linearity equation:
\begin{equation}\label{eq:linearity}
    \mathbf{U} a_j^\dagger \mathbf{U}^\dagger = \sum_{i=1}^N U^{1}_{ij} a_i^\dagger,
\end{equation}
where the $U^{1}_{ij}$ are the matrix elements of the $N \times N$ single-particle complex unitary matrix $U^{1}$.

\vspace{0.5cm}
\noindent \textbf{Proposition.} \textit{The transition amplitude of the multi-particle operator $\mathbf{U}$ between an initial $K$-particle occupation number representation state $\ket{\mathbf{m}}$ (associated with the index subset $J$) and a final $K$-particle state $\ket{\mathbf{l}}$ (associated with $I$) evaluates to the determinant of the $K \times K$ sub-matrix $U^{1}_{I,J}$ \cite{terhal2002classical, bravyi2002fermionic}:}
\begin{equation}
    \bra{\mathbf{l}} \mathbf{U} \ket{\mathbf{m}} = \det(U^{1}_{I, J}).
\end{equation}

\noindent \textbf{Proof.} Let the initial subset be $J = \{j_1 < j_2 < \dots < j_K\}$. We express the initial state in terms of $a^\dagger$ and insert the identity $\mathbb{I} = \mathbf{U}^\dagger \mathbf{U}$ between every operator:
\begin{equation}
\begin{aligned}
    \mathbf{U} \ket{\mathbf{m}} &= \mathbf{U} a_{j_1}^\dagger a_{j_2}^\dagger \dots a_{j_K}^\dagger \ket{\text{vac}} \\
    &= (\mathbf{U} a_{j_1}^\dagger \mathbf{U}^\dagger) (\mathbf{U} a_{j_2}^\dagger \mathbf{U}^\dagger) \dots (\mathbf{U} a_{j_K}^\dagger \mathbf{U}^\dagger) \mathbf{U} \ket{\text{vac}}.
\end{aligned}
\end{equation}
Using $\mathbf{U}\ket{\text{vac}} = \ket{\text{vac}}$ and Eq.~\eqref{eq:linearity}, substituting the linear transformation for each operator, we obtain:
\begin{equation}
    \mathbf{U} \ket{\mathbf{m}} = \left( \sum_{k_1=1}^N U^{1}_{k_1, j_1} a_{k_1}^\dagger \right) \dots \left( \sum_{k_K=1}^N U^{1}_{k_K, j_K} a_{k_K}^\dagger \right) \ket{\text{vac}}.
\end{equation}
We now project this state onto the final state $\bra{\mathbf{l}}$. Because states with different fermion occupation numbers are orthogonal, the inner product is non-zero if and only if $\{k_1, \dots, k_K\}$ is equal to $I = \{i_1, \dots, i_K\}$ as sets.

Therefore, the non-vanishing terms in the sum correspond exactly to the permutations $\sigma \in S_K$ of the list $(i_k)_k$, such that $k_m = i_{\sigma(m)}$. Due to the fermionic anticommutation relations, reordering the creation operators to the canonical ascending order yields the sign of the permutation:
\begin{equation}
    a_{i_{\sigma(1)}}^\dagger a_{i_{\sigma(2)}}^\dagger \dots a_{i_{\sigma(K)}}^\dagger \ket{\text{vac}} = \text{sgn}(\sigma) \ket{\mathbf{l}}.
\end{equation}
Factoring out the matrix elements and the permutation sign, the transition amplitude becomes:
\begin{equation}
    \bra{\mathbf{l}} \mathbf{U} \ket{\mathbf{m}} = \sum_{\sigma \in S_K} \text{sgn}(\sigma) \prod_{n=1}^K U^{1}_{i_{\sigma(n)}, j_n}.
\end{equation}
This is precisely Leibniz's formula for the determinant of the $K \times K$ matrix formed by the rows indexed by $I$ and columns indexed by $J$ of $U^{1}$. Thus, $\bra{\mathbf{l}} \mathbf{U} \ket{\mathbf{m}} = \det(U^{1}_{I,J})$. \hfill $\square$

\vspace{0.5cm}
\noindent\textbf{The $K$-particle sector and the compound matrix.} \\
Having established the transition amplitudes between individual multi-particle states, let us now characterize the overall block structure of the evolution operator, where each block is the $K$th compound matrix, i.e.\ dealing with the $K$-particle sector. The $\binom{N}{K}$ occupation number representation states $\ket{\mathbf{l}}$, labelled by $K$-element subsets $I \subseteq \{1,\dots,N\}$ and ordered lexicographically, form an orthonormal basis for the $K$-particle sector $\mathcal{H}_K$. By the Proposition, the matrix elements of $\mathbf{U}$ restricted to $\mathcal{H}_K$ are given by:
\begin{equation}
    \left(U^{K}\right)_{I,J} \equiv \bra{\mathbf{l}}\mathbf{U}\ket{\mathbf{m}} = \det(U^{1}_{I,J}).
\end{equation}
The right-hand side is precisely the $(I,J)$ entry of the $K$-th compound matrix of $U^{1}$ \cite{horn2012matrix}, which is the matrix constructed from all $K\times K$ minors of $U^{1}$:
\begin{equation}
    U^{K} = \mathrm{Minors}[U^{1}, K].
\end{equation}
This yields a $\binom{N}{K}\times\binom{N}{K}$ matrix. The compound matrices satisfy the following multiplicative property
\begin{equation}
    \mathrm{Minors}[AB, K] = \mathrm{Minors}[A, K]\,\mathrm{Minors}[B, K],
\end{equation}
as a direct consequence of the Cauchy--Binet theorem. Moreover, we have
\begin{equation}
    \mathrm{Minors}[U^{1}, K]^\dagger = \mathrm{Minors}[(U^{1})^\dagger, K],
\end{equation}
since the $(I,J)$ entry of $\mathrm{Minors}[U^{1},K]^\dagger$ is $\det((U^{1})_{J,I})^*$, while the $(I,J)$ entry of $\mathrm{Minors}[(U^{1})^\dagger,K]$ is $\det\!\left(((U^{1})^\dagger)_{I,J}\right)$.
The unitarity of $U^{K}$ then follows immediately:
\begin{equation}
    \left(U^{K}\right)^\dagger U^{K} = \mathrm{Minors}[(U^{1})^\dagger, K]\,\mathrm{Minors}[U^{1}, K] = \mathrm{Minors}[(U^{1})^\dagger U^{1}, K] = \mathrm{Minors}[\mathbb{I}_N, K] = \mathbb{I}_{\binom{N}{K}}.
\end{equation}

Hence, in the occupation basis, we obtain the following exact block-diagonal decomposition of $\mathbf{U}$ across all particle sectors:
\begin{equation}
    \mathbf{U} = \bigoplus_{K=0}^{N} U^{K}.
\end{equation}

This mathematical framework can be directly applied to the operators used to build the $SU(2)$-gauged Dirac QW in Sec.~\ref{sec:unitarity}. Specifically, the transport and coin operators $T_{x}^{1}$ and $C_{x}^{1}$, which were initially defined to act on the zero- and one-particle sectors of a $4$-mode local Fock space, can be written in the occupation basis $\ket{\mathbf{m}}$, with $\mathbf{m} \in \{0,1\}^{4}$:
\begin{equation}
   \ket{\mathbf{m}} =  ({a_{x}^{+}}^{\dagger})^{m_{1}}   ({a_{x}^{-}}^{\dagger})^{m_{2}}   ({b_{x+1}^{+}}^{\dagger})^{m_{3}}   ({b_{x+1}^{-}}^{\dagger})^{m_{4}}  \ket{0000}.
\end{equation}
Meanwhile, the chirality-swap operator $S_{x}^{1}$ is an on-site operator acting on the local Fock space, spanned by the basis $\ket{\mathbf{l}}$, with $\mathbf{l} \in \{0,1\}^{4}$:
\begin{equation}
   \ket{\mathbf{l}} =  ({b_{x}^{+}}^{\dagger})^{l_{1}}   ({b_{x}^{-}}^{\dagger})^{l_{2}}   ({a_{x}^{+}}^{\dagger})^{l_{3}}   ({a_{x}^{-}}^{\dagger})^{l_{4}} \ket{0000}.
\end{equation}

In these bases, we construct the full multi-particle extensions $\mathbf{S}_{x}$, $\mathbf{T}_{x}$, and $\mathbf{C}_{x}$ by applying the compound matrix decomposition to their single-particle counterparts $S_{x}^{1}$, $T_{x}^{1}$, and $C_{x}^{1}$:
\begin{align}
      \mathbf{S}_{x} &= \bigoplus_{K=0}^{4} \mathrm{Minors}[S_{x}^{1}, K], \\
      \mathbf{T}_{x} &= \bigoplus_{K=0}^{4} \mathrm{Minors}[T_{x}^{1}, K], \\
      \mathbf{C}_{x} &= \bigoplus_{K=0}^{4} \mathrm{Minors}[C_{x}^{1}, K].
\end{align}
Only $T^{1}_{x}$ has entries that are operators rather than numbers. But $\hat M$ is multiplication by $g$ (Eq.~\eqref{eq:Mismultbyg}), so at each fixed $g$ the transport is the ordinary unitary matrix $T^{1}_{x}(g)=\bigl(\begin{smallmatrix} 0 & g^\dagger \\ g & 0\end{smallmatrix}\bigr)$, to which the above applies verbatim, and $\mathbf{T}_{x}$ is obtained by applying $\mathrm{Minors}[T^{1}_{x}(g), K]$ at each $g$. Equivalently, the entries of $T^{1}_{x}$ are pointwise multiplications, hence commute.

The same applies to gauge covariance. Writing $Q=g_{x+1}\oplus g_{x}$ for the action of $G$ on the one-particle colour space, and recalling that on the link $G$ substitutes $g\mapsto g_x^\dagger\,g\,g_{x+1}$ (Eq.~\eqref{eq:bitranslation}), Sec.~\ref{sec:covariance} says precisely that $Q\,T^{1}_{x}(g_x^\dagger\,g\,g_{x+1})\,Q^\dagger=T^{1}_{x}(g)$. Taking $\mathrm{Minors}[\,\cdot\,,K]$ of both sides at each $g$, multiplicativity gives $G\,\mathbf{T}_{x}\,G^\dagger=\mathbf{T}_{x}$ in every particle sector. Consistently, $\mathrm{Minors}[g_x,2]=\det(g_x)=1$, which is the action $\gamma\ket{1,1}=\ket{1,1}$ of Eq.~\eqref{eq:gammafermions}.

Although these block-diagonal forms provide clear physical intuition by sectorizing the state space according to particle occupancy, this basis is highly impractical for direct implementation in a quantum computer. To bridge this gap and prepare the system for practical simulation, we map the system into the computational basis, ordering the states via standard binary counting with the rightmost qubit acting as the least significant bit. Under this transformation, the multi-particle operators $\mathbf{S}_{x}$, $\mathbf{T}_{x}$, and $\mathbf{C}_{x}$ elegantly factorize into tensor products of local one- and two-qubit operators:
\begin{align}
      \mathbf{S}_{x} &= (\mathbb{I}_2 \otimes \mathbf{S} \otimes \mathbb{I}_2) (\mathbf{S} \otimes \mathbf{S} )(\mathbb{I}_2 \otimes \mathbf{S} \otimes \mathbb{I}_2), \\
      \mathbf{M}_{x:R}&=\mathbf{M}^{\dagger} \otimes \mathbb{I}_2\\
      \mathbf{M}_{x+1:L}&=\mathbb{I}_2 \otimes \mathbf{M}\\
      \mathbf{T}_{x} &= \mathbf{S}_{x}(\mathbf{M}_{x+1:L}\mathbf{M}_{x:R}), \\
      \mathbf{C}_{x} &= (\mathbb{I}_2 \otimes \mathbf{S} \otimes \mathbb{I}_2) (\mathbf{C} \otimes \mathbf{C} )(\mathbb{I}_2 \otimes \mathbf{S} \otimes \mathbb{I}_2),
\end{align}
where the constituent operators $\mathbf{M}$, $\mathbf{S}$, and $\mathbf{C}$ are defined as:
\begin{equation}
    \mathbf{M} = 1 \oplus \hat{M} \oplus 1, \quad \mathbf{S} = 1 \oplus S \oplus -1, \quad \mathbf{C} = 1 \oplus C \oplus 1.
\end{equation}
Notice that $\mathbf{M}_{x+1:L}$ and $\mathbf{M}_{x:R}$ commute because they are controlled pointwise multiplications of scalar wavefunctions, see Eq.~\eqref{eq:multiplySU2gaugelabelsbymn}.
Their circuit representations are given below.


\begin{figure}[H]
    \centering
    \renewcommand{\baselinestretch}{1.2}

    \tikzset{
        block2/.style={draw=blue!80!black, thick, fill=blue!5, rounded corners=2pt, minimum width=1.4cm, minimum height=0.7cm, font=\bfseries},
        block3/.style={draw=blue!80!black, thick, fill=blue!5, rounded corners=2pt, minimum width=2.4cm, minimum height=0.7cm, font=\bfseries},
        cblock2/.style={draw=orange!80!black, thick, fill=orange!5, rounded corners=2pt, minimum width=1.4cm, minimum height=0.7cm, font=\bfseries},
        mblock/.style={draw=violet!80!black, thick, fill=violet!5, rounded corners=2pt, minimum width=2.8cm, minimum height=0.6cm, font=\bfseries},
        cdot/.style={circle, fill=black, inner sep=0pt, minimum size=4.5pt}
    }

    \begin{tikzpicture}[x=0.9cm, y=1.0cm, baseline=(current bounding box.center)]
        \node[font=\bfseries] at (2.5, -1.2) {(a) $\mathbf{S}_x$ (On-Site)};

        \draw[->, thick, gray] (-0.2, 0) -- (-0.2, 5.5) node[above, text=black] {$t$};

        \begin{scope}[on background layer]
            \fill[green!10, rounded corners=4pt] (0.5, 0) rectangle (4.5, 5.2);
            \node[green!80!black, font=\bfseries\small] at (2.5, 5.5) {Site $x$};
        \end{scope}

        \foreach \x/\label in {1/k_{l}^{+}, 2/k_{l}^{-}, 3/k_{r}^{+}, 4/k_{r}^{-}} {
            \draw[thick] (\x, 0) node[below] {$\ket{\label}$} -- (\x, 5);
        }

        \node[block2] at (2.5, 1.2) {$\mathbf{S}$};

        \node[block2] at (1.5, 2.6) {$\mathbf{S}$};
        \node[block2] at (3.5, 2.6) {$\mathbf{S}$};

        \node[block2] at (2.5, 4.0) {$\mathbf{S}$};
    \end{tikzpicture}
    \hfill
    \begin{tikzpicture}[x=0.9cm, y=1.0cm, baseline=(current bounding box.center)]
        \node[font=\bfseries] at (3, -1.2) {(b) $\mathbf{C}_x$ (On-Site)};

        \begin{scope}[on background layer]
            \fill[green!10, rounded corners=4pt] (0.5, 0) rectangle (4.5, 5.2);
            \node[green!80!black, font=\bfseries\small] at (2.5, 5.5) {Site $x$};
        \end{scope}

        \foreach \x/\label in {1/k_{l}^{+}, 2/k_{l}^{-}, 3/k_{r}^{+}, 4/k_{r}^{-}} {
            \draw[thick] (\x, 0) node[below] {$\ket{\label}$} -- (\x, 5);
        }

        \node[block2] at (2.5, 1.2) {$\mathbf{S}$};

        \node[cblock2] at (1.5, 2.6) {$\mathbf{C}$};
        \node[cblock2] at (3.5, 2.6) {$\mathbf{C}$};

        \node[block2] at (2.5, 4.0) {$\mathbf{S}$};
    \end{tikzpicture}
    \hfill
    \begin{tikzpicture}[x=0.9cm, y=1.0cm, baseline=(current bounding box.center)]
        \node[font=\bfseries] at (3, -1.2) {(c) $\mathbf{T}_x$ (Gauge Coupling)};

        \begin{scope}[on background layer]
            \fill[green!5, rounded corners=4pt] (0.5, 0) rectangle (2.5, 6.2);
            \node[green!80!black, font=\bfseries\small] at (1.5, 6.5) {Site $x$};

            \fill[cyan!5, rounded corners=4pt] (3.5, 0) rectangle (5.5, 6.2);
            \node[cyan!80!black, font=\bfseries\small] at (4.5, 6.5) {Site $x+1$};
        \end{scope}

        \foreach \x/\label in {1/k_{r}^{+}, 2/k_{r}^{-}, 4/k_{l}^{+}, 5/k_{l}^{-}} {
            \draw[thick] (\x, 0) node[below] {$\ket{\label}$} -- (\x, 6);
        }
        \draw[thick, magenta, decorate, decoration={snake, amplitude=1mm, segment length=4mm}]
            (3, 0) node[below, font=\scriptsize, align=center] {Gauge\\Link} -- (3, 6);

       
        \node[mblock] at (2, 0.8) {$\mathbf{M}^{\dagger}$};


        \node[mblock] at (4, 1.6) {$\mathbf{M}$};

        \node[block3] at (3, 2.7) {$\mathbf{S}$};

        \node[block2] at (1.5, 4.0) {$\mathbf{S}$};
        \node[block2] at (4.5, 4.0) {$\mathbf{S}$};

        \node[block3] at (3, 5.3) {$\mathbf{S}$};
    \end{tikzpicture}

    \vspace{0.5cm}
    \caption{Circuit representations of the $SU(2)$-gauged Dirac QCA. Time flows upwards. $\mathbf{S}_x$ and $\mathbf{C}_x$ locally at each site, whereas $\mathbf{T}_x$ spans across sites. The coupling operators $\mathbf{M}^\dagger$ and $\mathbf{M}$ are multi-qubit controlled unitary gates.}
    \label{fig:arrighi_qca_circuits}
\end{figure}
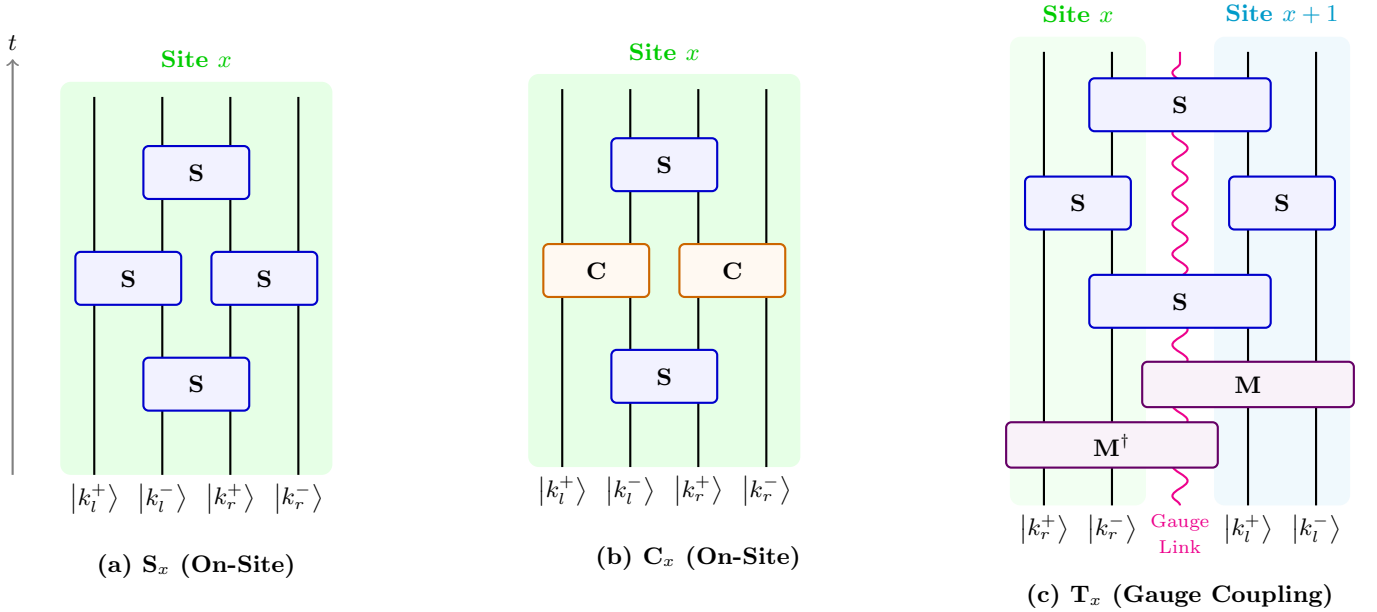

\section{Gauge link dynamics}\label{sec:interaction}

In $1+1$-dimensional lattice gauge theory, the pure gauge field dynamics are governed entirely by the electric field, as the lack of transverse spatial dimensions prohibits the existence of magnetic plaquette terms.

The local electrostatic energy of the gauge field is given by the sum of the squares of electric operators $\hat{E}^{a}$, $a=1,2,3$ and $\mathbf{E}^2 = \sum_{a} \hat{E}^{a}  \hat{E}^{a}$, which is mathematically identical to the quadratic Casimir operator of the $SU(2)$ Lie algebra. Recalling the Peter--Weyl decomposition of the link from Eq.~\eqref{eq:SU2gaugefield}, the orthogonal basis states $\ket{jm}\ket{jn}$ inherently diagonalize the Casimir operator. Its action depends exclusively on the total colour representation $j$:
\begin{align}
    \mathbf{E}^2\, \ket{jm}^{x:R}\ket{jn}^{(x+1):L} = j(j+1)\, \ket{jm}^{x:R}\ket{jn}^{(x+1):L}.\label{eq:casimir}
\end{align}

In the discrete-time framework of our QCA, this continuous Hamiltonian evolution is implemented by applying a local, diagonal unitary phase gate $U_E$ to each gauge link. Letting $\theta_{Y\!M} = \tfrac{1}{2}g_{Y\!M}^2 \Delta t \Delta x  = \tfrac{1}{2}g_{Y\!M}^2  \epsilon^{2}$ be a dimensionless parameter determined by the Yang--Mills coupling constant $g_{Y\!M}$ and the time-step size, the pure gauge evolution operator reads:
\begin{align}
    U_E(\theta_{Y\!M})\, \ket{jm}^{x:R}\ket{jn}^{(x+1):L} = \exp\left( -i \theta_{Y\!M} j(j+1) \right)\, \ket{jm}^{x:R}\ket{jn}^{(x+1):L}.\label{eq:UE}
\end{align}

Because the Casimir operator $\mathbf{E}^2$ is a multiple of the identity within each irreducible representation block $V_j$, it inherently commutes with the left and right actions of the group. Consequently, $U_E$ commutes with the local gauge transformations defined in Eqs.~\eqref{eq:halflinkR} and \eqref{eq:halflinkL}:
\begin{align}
    [U_E(\theta_{Y\!M}), G_x] = 0.
\end{align}
This guarantees that the addition of the electric field dynamics QCA sub-step independently preserves the exact $SU(2)$ gauge covariance of the system. The factor $\tfrac{1}{2}$ in $\theta_{Y\!M}$ is the one that makes $U_E$ the timestep of the electric Hamiltonian of Eq.~\eqref{eq:HE}: one power of $\epsilon$ becomes the timestep and the other the spatial measure.

\paragraph*{Truncation of the gauge link}
Because $\mathrm{e}^{-i\theta_{Y\!M} j(j+1)}$ depends on $\theta_{Y\!M} j(j+1)$ only modulo $2\pi$, we may consider possible truncation scenarios for the representations of the gauge links. By truncation, we mean imposing a ceiling on the gauge link representations, restricting them to $j = 0, 1/2, \dots, j_{\max}$. The value of $\theta_{Y\!M}$ determines the periodicity. We take it to be in the form of $\theta_{Y\!M} = 2 \pi / p$ for a rational number $p \in \mathbb{Q}$. We can capture the general periodic behavior by focusing on $p \in \mathbb{Z}^+$ and looking at
\begin{equation}
    U_{E}\left(\frac{2 \pi}{p}\right) \ket{j m} \ket{j n} = \exp\left[ -2 \pi i \left(\frac{j(j+1)}{p}\right) \right] \ket{j m} \ket{j n}.
\end{equation}
Since $j$ takes half-integer values, we substitute $j = (k-1)/2$ for $k \in \mathbb{N}^+$. The argument of the exponential becomes $-2 \pi i (k^2 - 1)/(4p)$. Consequently, the distinct physical phases belong to the sequence:
\begin{equation}
    (k^2 - 1) \pmod{4p}.
\end{equation}
Because this sequence is periodic, the phases come in a finite number and must therefore eventually repeat, thereby suggesting where to truncate the gauge link representations $j$ in the most harmless possible way. For example, when $p = 1$, the sequence $(k^2 - 1) \pmod{4}$ wraps up after $k = 2$, restricting the distinct representations to just $j = 0$ and $j = 1/2$. Based on this modular behavior, for an arbitrary integer $p$, the representations ought to be truncated at $j = p - 1/2$. One should note that in this truncation, there are still $k \leq p - 1/2$ that yield the same phase for $U_{E}$. This is because for each $k = p\pm t$, where $t < p$, we obtain the same phase. And when $p$ is a prime number, there are $p+1$ distinct phases. Additionally, there are further degeneracies when $p$ is not a prime. So that for an arbitrary $p$, there are at most $p+1$ distinct phases.

Naturally, a question arises: what happens beyond $j = j_{\max}$? There are two main possibilities. The first, and most naive approach, is to set the transition amplitudes to gauge link states with $j > j_{\max}$ to zero by modifying the Clebsch--Gordan coefficients that arise from the application of the operators $\hat{M}_{\pm, \pm}$. This procedure is inherently non-unitary: some probability is lost. Still, it makes the gauge-link Hilbert space finite and in terms of physical phenomenology we do not lose that much, as states with $j > j_{\max}$ tend to be energetically suppressed. 

The second approach is more elaborate; it alters the fundamental algebraic
structure of the theory so that the representation theory closes naturally
at $j_{\max}$, rather than imposing a harsh external truncation.  This relies on the so-called $q$-deformed $SU(2)$ quantum group \cite{Jimbo1985, Biedenharn1989, Macfarlane1989, Pasquier1990}. Instead of imposing a harsh, non-unitary cutoff, the theory is regularized by introducing a deformation parameter $q$ chosen as a root of unity. In this framework, standard integers within the Clebsch--Gordan coefficients are replaced by trigonometric quantum integers.  Interestingly, for the minimal truncation level $l=1$ (yielding $j_{\max}=1/2$), this $q$-deformed formulation exactly coincides with the $SU(2)$ Quantum Link Model \cite{Chandrasekharan1997}. In this specific limit, the modified algebraic rules map perfectly onto the physics of hardcore rishons (or hardcore Schwinger bosons). Because the hardcore constraint---akin to the Pauli exclusion principle---strictly forbids double-occupancy on the link, the state space naturally closes at $j=1/2$ without requiring the manual deletion of transition probabilities, a mathematical property successfully exploited in recent trapped-ion quantum simulation architectures \cite{Calajo2024}.\\

For our case, one might attempt a novel ``$q(p)$-deformed'' truncation scheme, drawing direct inspiration from the rich representation theory of $q$-deformed quantum groups at roots of unity.

\section{Gauss law}\label{sec:gauss-law}

A state $\ket{\Psi}$ is \emph{gauge-invariant} iff $G_x\ket{\Psi}=\ket{\Psi}$ for all $g_x\in SU(2)$ and all $x$ (Eqs.~\eqref{eq:G},\eqref{eq:overallGx}). At each site $x$, $G_x$ acts on three subsystems---the incoming half-link $x{:}L$, the fermions $\ket{\bold{k}_{r} \bold{k}_{l}}^x$, and the outgoing half-link $x{:}R$---each carrying a representation of $SU(2)$:
\begin{itemize}
\item $x{:}L$ carries $\pi^{j^{x:L}}$ (Eq.~\eqref{eq:halflinkL}).
\item $x{:}R$ carries $(\pi^{j^{x:R}})^*$ (Eq.~\eqref{eq:halflinkR}), which for $SU(2)$ is equivalent to $\pi^{j^{x:R}}$ (Eq.~\eqref{eq:SU2pistar}).
\item Each chirality of fermions independently carries $\pi^0$ if colourless ($\bold{k}_{r,l}\in\{(0,0),(1,1)\}$) or $\pi^{1/2}$ if coloured ($\bold{k}_{r,l}\in\{(1,0),(0,1)\}$), cf.\ Eq.~\eqref{eq:gammafermions}.
\end{itemize}
Writing $(j^x,m)$ for the combined colour representation and magnetic index, and $s^x$ for the spectator (transparent to $G_x$), recall that $G_x$ acts independently on the two chiralities (Eq.~\eqref{eq:overallGx}), leaving a colourless chirality untouched and acting as $g_x$ on the colour index of a coloured one. Explicitly:
\begin{enumerate}
\item For $j^x = 0$, $m = 0$: five spectators
\[
\ket{\mathbf{k}_l \mathbf{k}_r} \in \{\ket{1100},\; \ket{0011},\; \ket{0000},\; \ket{1111},\; \tfrac{1}{\sqrt{2}}(\ket{0110}-\ket{1001})\}.
\]
\item For $j^x = \tfrac{1}{2}$, $m = \pm\tfrac{1}{2}$: four spectators each,
\[
m = +\tfrac{1}{2}: \quad \{\ket{1000},\; \ket{0010},\; \ket{1011},\; \ket{1110}\}, \qquad
m = -\tfrac{1}{2}: \quad \{\ket{0100},\; \ket{0001},\; \ket{0111},\; \ket{1101}\}.
\]
\item For $j^x = 1$: one spectator, namely $\ket{1010}$ for $m=+1$, $\tfrac{1}{\sqrt{2}}(\ket{0110}+\ket{1001})$ for $m=0$, and $\ket{0101}$ for $m=-1$.
\end{enumerate}
So the multiplicities of the spectator are $j^x=0\to 5$, $j^x=\tfrac12\to 4$ and $j^x=1\to 1$. Multiplying by the number of allowed $m$ for each $j^x$, this checks out since $5\times1+4\times2+1\times3=16=2^4$.
The only case worth spelling out is when both chiralities are coloured, i.e.\ the last state of the first list together with the whole of the third. There $G_x$ acts as $g_x$ on the colour index of the right-mover and again as $g_x$ on that of the left-mover, so the combination $\tfrac{1}{\sqrt{2}}(\ket{0110}-\ket{1001})$ merely picks up $\det(g_x)=1$---the same mechanism as in Eq.~\eqref{eq:detM}. It is thus left invariant by $G_x$ and provides a fifth spectator for $j^x=0$, the three remaining combinations being those listed for $j^x=1$.

At site $x$, $G_x$ acts as $\pi^{j^{x:L}}(g_x)\otimes\pi^{j^x}(g_x)\otimes\pi^{j^{x:R}}(g_x)^*$ on the three subsystems (Eqs.~\eqref{eq:halflinkL},\eqref{eq:gammafermions},\eqref{eq:halflinkR}). Since $\pi^{j*}\cong\pi^j$ for $SU(2)$ (Eq.~\eqref{eq:SU2pistar}), gauge invariance requires finding the trivial component of $\pi^{j^{x:L}}\otimes\pi^{j^x}\otimes\pi^{j^{x:R}}$. By the Clebsch--Gordan theorem (Eq.~\eqref{eq:CG-decomposition}), $\pi^{j^x}\otimes\pi^{j^{x:R}}\cong\bigoplus_{k=|j^x-j^{x:R}|}^{j^x+j^{x:R}}\pi^k$, where $k$ advances in unit steps. Tensoring with $\pi^{j^{x:L}}$, the trivial representation $\pi^0$ appears iff $k=j^{x:L}$ lies in this range, i.e.\ $|j^x{-}j^{x:R}|\leq j^{x:L}\leq j^x{+}j^{x:R}$ together with $j^{x:L}{+}j^x{+}j^{x:R}\in\mathbb{Z}$, which is what places $j^{x:L}$ on the ladder and not merely between its endpoints. The sectors $\pi^k$ with $k>0$ are non-trivial representations, hence no state in them is invariant under all $g_x$; only the one-dimensional $\pi^0$ sector survives, fixing the colour-index structure uniquely. This condition is equivalent to
\begin{align}
|j^{x:L}-j^{x:R}|\leq j^x\leq j^{x:L}+j^{x:R},&\qquad j^{x:L}+j^x+j^{x:R}\in\mathbb{Z},\label{eq:triangle}\\
|\Delta j^{x:R}|\leq j^{x+1}\leq j^{x:R}+j^{x+1:R},&\qquad j^{x:R}+j^{x+1}+j^{x+1:R}\in\mathbb{Z},\label{eq:triangleB}
\end{align}
with $\Delta j^{x:R}:=j^{x+1:R}-j^{x:R}$ and using $j^{x:R} = j^{x+1:L}$.
This is the non-abelian discrete Gauss law: the colour charge $j^x$ at site $x$ is constrained by the adjacent link representations. Notice that for half-integer labels the inequalities alone would not suffice: e.g.\ $j^{x:L}=j^x=j^{x:R}=\tfrac12$ satisfies them, and yet $\pi^{1/2}\otimes\pi^{1/2}\cong\pi^0\oplus\pi^1$ admits no $\pi^{1/2}$.

\noindent{\em Physical Hilbert space.}
Let $\ket{j^x,\,m,\,s^x}^x$ denote the four-qubit fermion state with colour content $(j^x,m)$ and spectator $s^x$ as above. By Eq.~\eqref{eq:SU2gaugefield} each link is shared between adjacent sites. To construct the gauge-invariant state at site $x$, couple the fermion and the incoming half-link $x{:}L$ via Eq.~\eqref{eq:CG-expansion}:
\begin{align}
\ket{j^x,\,m}^x\,\ket{j^{x:L},\,m'}^{x:L} \cong \sum_{k} \braket{j^x\,m,\;j^{x:L}\,m'}{k\,(m+m')}\;\ket{k,\,m+m'}^{x,x:L}.\label{eq:couplexxL}
\end{align}
Under $G_x$ the coupled ket transforms as $\pi^k(g_x)$, while $\ket{j^{x:R},\,n'}^{x:R}$ transforms as $\pi^{j^{x:R}}(g_x)^*$ (Eq.~\eqref{eq:halflinkR}). By unitarity of $\pi^j(g)$, the unique state invariant under $\pi^j(g)^*\otimes\pi^j(g)$ for all $g$ is:
\begin{align}
\sum_{q}\ket{j,\,q}^{x:R}\otimes\ket{j,\,q}^{x,x:L}
\end{align}
which forces $k=j^{x:R}$. Looking at who can generate $\ket{j^{x:R},\,q}^{x,x:L}$ via Eq.~\eqref{eq:CG-expansion} we get:
\begin{align}
\ket{\Phi_x^{j^x,s^x}} \;\propto\; &\sum_{m,\, m'} \braket{j^x\,m,\; j^{x:L}\,m'}{j^{x:R}\,(m{+}m')}\;\ket{j^{x:R},\,m{+}m'}^{x:R}\;\ket{j^x,\,m,\,s^x}^x\;\ket{j^{x:L},\,m'}^{x:L},\label{eq:localfactor}
\end{align}
The spectator $s^x$ is free.
The general gauge-invariant state is thus
\begin{align}
\ket{\Psi} = \sum_{\mathbf{J}} \alpha_{\mathbf{J}} \bigotimes_x \ket{\Phi_x^{j^x,\,s^x}},\label{eq:physicalstate}
\end{align}
where $\mathbf{J}$ ranges over possible representation assignments each of the form $(j^{x:L}, j^x,j^{x:R},s^x)_x$, each satisfying both the triangle condition~\eqref{eq:triangle}, and that $s^x$ is within the multiplicity allowed by $j^x$, at every $x$.

\noindent{\em Superselection sectors.}
As in the $U(1)$ case, the physical Hilbert space decomposes into superselection sectors. The total fermion number $N=\sum_x\bigl(|\mathbf{k}_l|^x+|\mathbf{k}_r|^x\bigr)$ is conserved by the QCA dynamics. There is no analogous per-link conservation, however: the comparator $\hat M$ (Eq.~\eqref{eq:multiplySU2gaugelabelsbymn}) shifts each link representation by $\pm\tfrac{1}{2}$, so a hop flips the parity of $2j^{x:R}$ on the link it crosses.

\noindent{\em Spin networks.}
Notice that in this $1{+}1$-dimensional theory, only the $j^x,j^{x:R},s^x$ labels are free: the superposition over magnetic indices $m,m'$ in each $\ket{\Phi_x^{j^x,s^x}}$ is entirely fixed by the Clebsch--Gordan coefficients in Eq.~\eqref{eq:localfactor}. One could therefore work entirely in the reduced physical Hilbert space labeled by $(j^x,j^{x:R},s^x)_x$, eliminating the magnetic indices---this is the spin network basis \cite{RovelliSmolin1995Spin}. We retain the full $L^2(SU(2))$ description because the comparator $\hat M$ (Eq.~\eqref{eq:multiplySU2gaugelabelsbymn}) and the QCA sub-steps (Sec.~\ref{sec:unitarity}) are naturally expressed in terms of mode labels; in the spin network basis, the same dynamics would require $6j$-symbols (Racah--Wigner recoupling coefficients) and the local tensor-product structure of the QCA would become less transparent. Moreover, it is this description that generalizes to higher dimensions.

\section{Altogether}\label{sec:altogether}

Combining the results of Secs.~\ref{sec:unitarity}--\ref{sec:gauss-law},
we now assemble the complete $SU(2)$ QCA.

\noindent{\em Complete one-timestep evolution.}
The full $SU(2)$ QCA timestep $\mathbf{U}$ is the sequential composition
of the $SU(2)$-gauged Dirac QCA $\mathbf{W}$ (Sec.~\ref{sec:multi-particle})
and the pure-gauge electric sub-step $U_E$ (Sec.~\ref{sec:interaction}):
\begin{align}
    \mathbf{U} \;=\; U_E(\theta_{Y\!M})\cdot\mathbf{W},
    \label{eq:QCAcomplete}
\end{align}
where $\theta_{Y\!M}$ is as defined in Sec.~\ref{sec:interaction}, and the
$SU(2)$-gauged Dirac QCA decomposes as
\begin{align}
    \mathbf{W} \;=\; \mathbf{C}\cdot\mathbf{T}\cdot\mathbf{S}.
    \label{eq:Wdecomp}
\end{align}
Here $\mathbf{S}$, $\mathbf{T}$, $\mathbf{C}$ are the multi-particle
extensions (Sec.~\ref{sec:multi-particle}) of the chirality swap
(Eq.~\eqref{eq:SSU2}), cross-site transport (Eq.~\eqref{eq:TSU2}),
and mass coin (Eq.~\eqref{eq:CSU2}), respectively, and $U_E$ is the
diagonal phase gate of Eq.~\eqref{eq:UE} applied to every gauge link.
In $1+1$ dimensions, the absence of transverse spatial directions
means there is no magnetic plaquette term; the gauge-field dynamics
reduces entirely to $U_E$.

\noindent{\em Global assembly from local gates.}
The global operators in Eqs.~\eqref{eq:QCAcomplete}--\eqref{eq:Wdecomp}
are tensor products of the per-site operators defined in
Secs.~\ref{sec:unitarity}--\ref{sec:multi-particle}:
\begin{align}
    \mathbf{S} = \bigotimes_x \mathbf{S}_x, \qquad
    \mathbf{C} = \bigotimes_x \mathbf{C}_x, \qquad
    \mathbf{T} = \bigotimes_x \mathbf{T}_x, \qquad
    U_E(\theta_{Y\!M}) = \bigotimes_x U_E^{x,x+1}(\theta_{Y\!M}),
    \label{eq:globalops}
\end{align}
where $\mathbf{S}_x$, $\mathbf{C}_x$ (Eqs.~\eqref{eq:SSU2},\eqref{eq:CSU2})
are strictly on-site, $\mathbf{T}_x$ (Eq.~\eqref{eq:TSU2}) spans sites $x$,
$x+1$ and the link between them, and $U_E^{x,x+1}$ (Eq.~\eqref{eq:UE}) acts
on that link alone. Since $\mathbf{T}_x$ and $\mathbf{T}_{x+1}$ have disjoint
supports (see Fig.~\ref{fig:arrighi_qca_circuits}), $\bigotimes_x \mathbf{T}_x$ is well-defined and applied simultaneously
across all links.

\noindent{\em Gauge covariance.}
$\mathbf{S}$ and $\mathbf{C}$ act as the identity on colour and on the
link; $\mathbf{T}$ is gauge-covariant by the cancellation
$g_x^\dagger g_x = g_{x+1}^\dagger g_{x+1} = \mathbb{I}_2$ of
Sec.~\ref{sec:covariance}, extended to all particle sectors in
Sec.~\ref{sec:multi-particle}; and $U_E$ is gauge-covariant because the
Casimir $\mathbf{E}^2$ commutes with every local gauge transformation
(Sec.~\ref{sec:interaction}). Consequently,
\begin{align}
    \bigl[G,\,\mathbf{U}\bigr]\;=\;0.\label{eq:QCAgaugeinv}
\end{align}

\noindent{\em Unitarity.}
$\mathbf{S}$ and $\mathbf{C}$ are manifestly unitary; $\mathbf{T}$ is
unitary by Eq.~\eqref{eq:MMdagger}, whose proof relied on $\det(g)=1$
for $g\in SU(2)$; and $U_E$ is a diagonal phase gate. Their composition
$\mathbf{U}$ is therefore unitary.

\noindent{\em Physical Hilbert space and Gauss law.}
Since $\mathbf{U}$ is gauge-covariant (Eq.~\eqref{eq:QCAgaugeinv}), it
preserves the physical subspace $\mathcal{H}_{\mathrm{phys}}$:
\begin{align}
    \forall\;\ket{\Psi}\in\mathcal{H}_{\mathrm{phys}},\qquad
    \mathbf{U}\ket{\Psi}\in\mathcal{H}_{\mathrm{phys}},
\end{align}
where gauge-invariant states take the form of Eq.~\eqref{eq:physicalstate}
and satisfy the Gauss law of Eqs.~\eqref{eq:triangle}--\eqref{eq:triangleB}.\\

\noindent{\em Continuum limit.}
Let $\epsilon=\Delta t=\Delta x$ denote the lattice spacing, with sites at $x\in\epsilon\mathbb{Z}\subset\mathbb{R}$. We now sketch the continuum limit ($\epsilon \to 0$) of the fermionic and gauge-field Hilbert spaces that make up the QCA $\mathbf{W}$. This is a sketch in the following sense: we expand each gate to first order in $\epsilon$ and read off the generator, but we do not prove that these expansions converge. Turning this into a theorem---providing the embeddings, the operator domains and the topology in which the limits are taken---is left to future work. The expansion of the comparator below, for instance, is a pointwise one, valid near the identity of $SU(2)$; it is not a statement about $\hat{M}$ as an operator on the whole of $L^2(SU(2))$.

\emph{Fermionic sector.} For a fixed chirality and colour, the discrete one-particle position space $\ell^2(\epsilon\mathbb{Z})$ transitions into the continuous space $L^2(\mathbb{R})$. Embedding discrete states $\psi\in\ell^2(\epsilon\mathbb{Z})$ as, e.g., functions, the discrete norm converges to the standard integral $\int_{\mathbb{R}}dx\,|\psi(x)|^2$, and the lattice shift operator converges to the continuous translation operator $e^{\epsilon\partial_x}$ \cite{ArrighiNesmeForets2014}. Thus, the one-particle state space becomes:
\begin{align}
\ell^2(\epsilon\mathbb{Z})\otimes\mathbb{C}^2_{\mathrm{chir}}\otimes\mathbb{C}^2_{\mathrm{col}}\ \xrightarrow{\ \epsilon\to0\ }\ L^2(\mathbb{R})\otimes\mathbb{C}^2_{\mathrm{chir}}\otimes\mathbb{C}^2_{\mathrm{col}},
\label{eq:matterlimit}
\end{align}
yielding four-component continuum wavefunctions $\Psi(x)=(\psi_r,\psi_l)^T(x)$, where $\psi_r,\psi_l\in\mathbb{C}^2$.

\emph{Gauge sector.} Each lattice link carries a state in $L^2(SU(2))$ with $dg$, the normalized Haar measure. The $SU(2)$ group manifold is compact and curved; however, near its identity matrix $\mathbb{I}_2$, it is locally flat and is mapped directly to its tangent space, the Lie algebra $\mathfrak{su}(2) \cong \mathbb{R}^3$, referred to as the set of connections. Consider $A_1\in\mathbb{R}^3$ as such a connection, we can rescale the local coordinates as $\theta^a=\epsilon\,g_{Y\!M}\,A_1^a$ with $a=1\ldots 3$. As $\epsilon \to 0$, the link variables shrink into an $O(\epsilon)$ neighbourhood of $\mathbb{I}_2$. In this neighbourhood, the Haar measure flattens into the standard Lebesgue measure, $dg \to d^3A_1$ \cite{Faraut2008,Hall2015}.


Consequently, just as a compact $U(1)$ circle unwraps locally onto the real line in Abelian lattice gauge theory \cite{ArrighiQED}, the configuration space $SU(2)$ admits an In\"on\"u--Wigner contraction \cite{InonuWigner1953, Thiemann2007} to its flat tangent space near the identity. Correspondingly,
\begin{align}
L^2(SU(2)) \,\, \textit{with} \,\, dg\ \xrightarrow{\ \epsilon\to0\ }\ L^2(\mathbb{R}^3) \,\, \textit{with} \,\,  d^3A_1.
\label{eq:linklimit}
\end{align}
This aligns with the standard correspondence in lattice gauge theory \cite{KogutSusskind1975}. Globally, the continuous gauge sector becomes a flat connection-representation space. While heuristically written as the continuous infinite tensor product $\bigotimes_{x\in\mathbb{R}}L^2(\mathbb{R}^3)_x$, this object is mathematically ill-defined due to the non-existence of a translation-invariant infinite-dimensional Lebesgue measure \cite{Oxtoby1946}. One possible route towards making it rigorous is to smear the operator $\hat{A}_1^a(x)$ over the position space. Along that route, one can replace $\bigotimes_{x\in\mathbb{R}}L^2(\mathbb{R}^3)_x$ with $L^2(\mathcal{S}'(\mathbb{R}, \mathfrak{su}(2))) \simeq L^2(\mathcal{S}'(\mathbb{R},\mathbb{R}^3))$ with a Gaussian measure $d\mu_C$ via the Bochner--Minlos theorem \cite{GlimmJaffe1987}---though that theorem only delivers such a measure once a covariance $C$ has been chosen, and we do not identify here which choice, if any, arises as the limit of the lattice Haar measures.

\emph{Covariant derivative and fermionic Hamiltonian.} Most of the authors who work at the level of operators, not algebras, consider the Taylor expansions of the parallel transporter to get to the gauge field $\hat{A}_1^a(x)$ \cite{KogutSusskind1975}. If we were to do this to the comparator operator $\hat{M}_{x,x+1}$, it expands as:
\begin{align}
\hat{M}_{x,x+1} = \mathbb{I}_2 + i\epsilon\,g_{Y\!M}\,\hat{A}_1^a(x)\,\frac{\sigma^a}{2} + O(\epsilon^2).
\label{eq:Mexpand}
\end{align}
Composing this with the spatial shift yields the gauge-covariant derivative: $e^{\epsilon\partial_x}\hat{M}_{x,x+1} = \mathbb{I}_2 + \epsilon\,D_x + O(\epsilon^2)$, where
\begin{align}
D_x := \partial_x + i\,g_{Y\!M}\,\hat{A}_1^a(x)\,\frac{\sigma^a}{2},
\end{align}
acting on $L^2(\mathbb{R})\otimes\mathbb{C}^2_{\mathrm{col}}\otimes L^2(\mathbb{R}^3)_x$. By Sec.~\ref{sec:multi-particle}, $\mathbf{W}$ is the compound-matrix extension, sector by sector in particle number, of the single-particle unitary $W^1=C^1T^1S^1$; its fermionic continuum limit is therefore fixed entirely by that of $W^1$. Expanding $W^1$ to first order in $\epsilon$ gives the fermionic Hamiltonian $H_F$:
\begin{align}
H_F = -i\,(\sigma_z\otimes\mathbb{I}_2)\,D_x + m\,(\sigma_x\otimes\mathbb{I}_2),
\label{eq:HF}
\end{align}
where $\sigma_z,\sigma_x$ act on chirality and $\mathbb{I}_2$ acts on colour. This $H_F$ is the one-particle generator; on the many-particle space it is its second quantisation that plays this role, i.e.\ the corresponding integral of field bilinears. When $g_{Y\!M}=0$, $H_{F}$ correctly reduces to the free Dirac QCA \cite{ArrighiNesmeForets2014}. This Hamiltonian matches well with the $SU(2)$ limit of the Hamiltonian operators at Ref.~\cite{PhysRevD.22.939}. Furthermore, because our continuum limit yields, at first order in $\epsilon$, the $1+1$-dimensional $SU(2)$ Yang--Mills theory coupled to a massive Dirac fermion, this QCA can be viewed as a discrete, Hamiltonian lattice regularisation of the massive gauged Wess--Zumino--Witten (WZW) model \cite{Witten1984}.\\

\emph{Pure-gauge sector.} The purely gauge dynamics are governed by the unitary $U_E$, which applies a phase dependent on the electric field. The electric field operator $\hat{E}^a_{x}$ is the conjugate momentum to the $SU(2)$ link variable, and $\mathbf{E}^2$ is the quadratic Casimir operator acting as the Laplace--Beltrami operator on $SU(2)$. Under the continuum rescaling and moving to $\hat{M}= \mathbb{I}_2 + i \epsilon\, g_{Y\!M}\, \hat{A}_1^a \frac{\sigma^{a}}{2}$, the conjugate continuum electric field operator is $\hat{E}^a(x)$, which acts as the functional derivative $-i\delta / \delta \hat{A}_1^a(x)$ on the continuous limit space. Consequently, the Casimir operator flattens into the standard flat Laplacian on $\mathbb{R}^3$. Expanding the unitary as $U_E = \mathbb{I} - i\epsilon H_E + O(\epsilon^2)$ yields the continuous electric Hamiltonian:
\begin{align}
H_E = \int_{\mathbb{R}} dx \, \frac{g_{Y\!M}^2}{2} \big(\hat{E}_1^a(x)\big)^2.
\label{eq:HE}
\end{align}

\emph{Full continuum limit.} The complete QCA step composes the matter and pure-gauge updates (e.g., $\bold{U}= U_E \mathbf{W}$). Expanding this total timestep to first order in $\epsilon$ via $\bold{U} = \mathbb{I} - i\epsilon H + O(\epsilon^2)$ yields the full continuous Schr\"odinger equation $i\partial_t\Psi = H\Psi$, governed by the total Hamiltonian:
\begin{align}
H = H_F + H_E = -i\,(\sigma_z\otimes\mathbb{I}_2)\,D_x + m\,(\sigma_x\otimes\mathbb{I}_2) + \int_{\mathbb{R}} dx \, \frac{g_{Y\!M}^2}{2} \big(\hat{E}_1^a(x)\big)^2.
\label{eq:Hfull}
\end{align}
This is the expected Hamiltonian of $1+1$-dimensional $SU(2)$ Yang--Mills theory minimally coupled to a Dirac fermion, matching the continuous limit of $2D$ non-Abelian quantum chromodynamics \cite{FrishmanSonnenschein1993}.

\section{Conclusion}\label{sec:conclusion}

{\em Summary of contributions.}

In this paper we constructed a quantum cellular automaton (QCA) realising $1+1$-dimensional $SU(2)$ Yang--Mills theory with Dirac fermions---to the authors' knowledge the first QCA for a non-abelian gauge theory.

We first constructed a gauge-covariant fermionic transfer. The naive hopping $T_R=\sum_{m,n} {b^n}^\dagger M^*_{mn}\,a^m$ \eqref{eq:TR} fails to commute with independent local $SU(2)$ transformations; this is fixed by promoting $M$ to an operator $\hat M$ acting on a gauge field living in $L^2(SU(2))$ \eqref{eq:SU2gaugefield}, defined not through Clebsch--Gordan sums but through the pointwise product of Wigner modes \eqref{eq:multiplySU2gaugelabelsbymn}. This structure permitted line-by-line proofs of its gauge covariance $G(\hat M)=g_x^\dagger\,\hat M\,g_{x+1}$ \eqref{eq:comptransf}, hence of the gauge covariance of the resulting transfer \eqref{eq:slippery}, together with its unitarity $\hat M^\dagger\hat M=\mathbb{I}_2$ \eqref{eq:MMdagger}.

We provided a self-contained summary of how to lift any such single-particle gate to the multi-particle sector whilst preserving both unitarity and gauge covariance. Namely, we explained how the non-interacting extension acts on the $K$-particle sector as the $K$th compound matrix \eqref{eq:linearity}, from which unitarity follows via the Cauchy--Binet theorem, and we exhibited the resulting decomposition into local one- and two-qubit gates (Fig.~\ref{fig:arrighi_qca_circuits}), yielding the multi-particle $SU(2)$-gauged Dirac QCA $\mathbf{W}=\mathbf{C}\cdot\mathbf{T}\cdot\mathbf{S}$ \eqref{eq:Wdecomp}.

We then turned on the gauge-field dynamics: the electric energy term is implemented by a diagonal phase gate $U_E$ \eqref{eq:UE}. We also derived the gauge-invariant subspace: only at this final stage---where unitarity is no longer at stake---did we invoke Clebsch--Gordan theory, to derive the non-abelian discrete Gauss law \eqref{eq:triangle}.

Finally, we sketched the continuum limit $\varepsilon\to0$, in which the QCA yields, at first order in $\varepsilon$, the expected Hamiltonian of $1+1$-dimensional $SU(2)$ Yang--Mills theory minimally coupled to a massive Dirac fermion \eqref{eq:Hfull}; a convergence proof is left to future work.

Two methodological threads run through all of the above: a systematic reliance on the pointwise-product structure of the gauge-link updates in place of Clebsch--Gordan sums, which is what made the constructive, gate-by-gate proofs of unitarity and covariance possible; and a deliberate recasting of the traditional quantum-field-theory construction into modern quantum-information vocabulary and notation.

\noindent {\em Perspectives.}\\

A first perspective concerns the gauge link itself. Being an instance of $L^2(SU(2))$ \eqref{eq:SU2gaugefield}, it carries infinitely many representations $j$, which any actual implementation must truncate. We observed that the $2\pi$-periodicity of the electric phase gate $U_E$ \eqref{eq:UE} singles out a finite set of distinct phases and suggests a natural, unitary cutoff through a novel ``$q(p)$-deformation'' of $SU(2)$, in the spirit of the quantum-group regularisations underlying quantum link models.

A second perspective is to raise the spatial dimension. The $1+1$ construction is deliberately minimal---in particular, the absence of transverse directions means there is no magnetic plaquette term, and the gauge-field dynamics reduces to the electric phase gate $U_E$ \eqref{eq:UE}. Extending the QCA to $2+1$ and $3+1$ dimensions would require reinstating the magnetic contribution and enlarging the fermionic register, in analogy with the abelian $3+1$ QED QCA \cite{Eon_2023}.

A third perspective, and our best hope for reaching general $SU(N)$, is to decompose the comparator $\hat M$ \eqref{eq:multiplySU2gaugelabelsbymn}---the only genuinely gauge-charged gate of the circuit---into still smaller primitive gates. Such a decomposition, obtainable through a Schur-decomposition algorithm \cite{Bacon_2006}, would additionally open the way to a purely graphical proof of gauge covariance.

A fourth perspective is to handle the Fermion Doubling problem, i.e., the persisting spurious solutions at any $\varepsilon$, however small---by applying the method given in \cite{bakircioglu2026fermiondoublingquantumcellular,bakircioglu2026flavouredlatticeschwingermodel}. This will likely motivate new flavour degrees of freedom, in a way that is reminiscent of the Standard Model.

\subsection*{Acknowledgements}

The authors would like to deeply thank Giuseppe Magnifico for receiving Dogukan Bakircioglu and teaching him the basics of $SU(2)$ lattice gauge theories. They wish to thank Pablo Arnault for his early insights.
This project was partially funded by the European Union through the MSCA SE project QCOMICAL, by the French National Research Agency (ANR): projects TaQC ANR-22-CE47-0012 and within the framework of `Plan France 2030', under the research projects EPIQ ANR-22-PETQ-0007, OQULUS ANR-23-PETQ-0013, HQI-Acquisition ANR-22-PNCQ-0001 and HQI-R\&D
ANR-22-PNCQ-0002, and by the WOST, WithOut SpaceTime project (https://withoutspacetime.org), grant ID\# 63683 from the John Templeton Foundation (JTF). The opinions expressed in this work are those of the author(s) and do not necessarily reflect the views of the John Templeton Foundation.

\bibliographystyle{quantum}
\bibliography{main}

\appendix

\section{The Peter-Weyl theorem}\label{app:peter-weyl}

We recall the Peter-Weyl theorem \cite{Sepanski2006-mh}, adapted to our conventions.
Let $G$ be a compact group with normalized Haar measure ($\int_G dg = 1$). Let $\hat{G}$ denote a complete set of pairwise inequivalent irreducible unitary representations of $G$. For each $\pi^j \in \hat{G}$, let $V_j$ be its representation space, $d_j := \dim V_j$, and $\{\ket{j\mu}\}_{\mu=1}^{d_j}$ an orthonormal basis of $V_j$. Define the matrix coefficients $D^j_{\mu\nu}(g) := \bra{j\mu}\pi^j(g)\ket{j\nu}$. We use the left/right translation convention $(\Lambda_{h_L,h_R}\psi)(g):=\psi(h_L\,g\,h_R^{-1})$.

\noindent\textbf{Theorem} (Peter-Weyl).
\begin{enumerate}
\item[(i)] Every unitary representation of $G$ on a Hilbert space decomposes as an orthogonal direct sum of irreducible finite-dimensional unitary representations.
\item[(ii)] As a representation of $G \times G$ acting by $\Lambda$,

\begin{align}
    L^2(G) \cong \bigoplus_{j \in \hat{G}} V_j^* \otimes V_j.\label{eq:PW-decomposition}
\end{align}
\item[(iii)] The functions $\{\sqrt{d_j}\, D^j_{\mu\nu}\}$ form an orthonormal basis of $L^2(G)$:
\begin{align}
    \int_G dg\, (D^{j'}_{\mu'\nu'}(g))^* D^j_{\mu\nu}(g) = \frac{\delta_{jj'}\delta_{\mu\mu'}\delta_{\nu\nu'}}{d_j}.\label{eq:PW-orthogonality}
\end{align}
\end{enumerate}

\noindent In the main text we apply this to $G=U(1)$ (with $\hat{G}=\mathbb{Z}$, $d_l=1$, $D^l(g)=g^l$) and $G=SU(2)$ (with $j=0,\tfrac12,1,\ldots$, $m,n\in\{-j\ldots j\}$, $d_j=2j+1$, and $D^j_{mn}(g)=\bra{jm}\pi^j(g)\ket{jn}$).

\section{Clebsch--Gordan coefficients and the pointwise product}\label{app:CG}

The tensor product of two irreducible representations of $SU(2)$ decomposes as a direct sum of irreducible representations:
\begin{align}
    \pi^j\otimes\pi^{j'} \cong \bigoplus_{k=|j-j'|}^{j+j'} \pi^k.\label{eq:CG-decomposition}
\end{align}
Concretely, there exists a unitary change-of-basis $C: V_j\otimes V_{j'}\to \bigoplus_{k} V_k$ satisfying
\begin{align}
    C\,\bigl(\pi^j(g)\otimes\pi^{j'}(g)\bigr)\,C^\dagger = \bigoplus_{k=|j-j'|}^{j+j'} \pi^k(g) \qquad\forall g\in SU(2),\label{eq:CG-intertwine}
\end{align}
The matrix elements of $C$ in the uncoupled and coupled bases are given by
\begin{align}
    C\ket{jm,j'n} \cong \sum_{k=|j-j'|}^{j+j'} \braket{jm,j'n}{k\,(m\!+\!n)}\;\ket{k\,(m\!+\!n)},\label{eq:CG-expansion}
\end{align}
These are the Clebsch--Gordan coefficients $\braket{jm,j'n}{k\,q}\in\mathbb{R}$. The selection rule $q=m+n$ follows from the fact that $C$ commutes with $S_z^{\mathrm{tot}} = S_z\otimes\mathbb{I}+\mathbb{I}\otimes S_z$.

\medskip
\noindent We now derive Eq.~\eqref{eq:multiplySU2gaugemodeswithCG}. By definition of the Wigner $D$-matrix elements:
\begin{align}
    D^j_{mn}(g)\,D^{j'}_{m'n'}(g)
    &= \bra{jm}\pi^j(g)\ket{jn}\;\bra{j'm'}\pi^{j'}(g)\ket{j'n'}\label{eq:pw1}\\
    &= \bra{jm,j'm'}\;\pi^j(g)\otimes\pi^{j'}(g)\;\ket{jn,j'n'}.\label{eq:pw2}
\end{align}
Step~\eqref{eq:pw2} uses the standard identity $\bra{\alpha}A\ket{\beta}\,\bra{\alpha'}B\ket{\beta'} = \bra{\alpha,\alpha'}(A\otimes B)\ket{\beta,\beta'}$. Inserting $\mathbb{I} = C^\dagger C$ on both sides of the operator and applying~\eqref{eq:CG-intertwine}:
\begin{align}
    &= \bra{jm,j'm'}C^\dagger\;\Bigl(\bigoplus_{k}\pi^k(g)\Bigr)\;C\ket{jn,j'n'}.\label{eq:pw3}
\end{align}
Expanding $C$ via~\eqref{eq:CG-expansion} on the ket, and $C^\dagger$ on the bra (using reality of CG coefficients):
\begin{align}
    &= \sum_{k=|j-j'|}^{j+j'} \braket{jm,j'm'}{k\,(m\!+\!m')}\;\underbrace{\bra{k\,(m\!+\!m')}\pi^k(g)\ket{k\,(n\!+\!n')}}_{D^k_{m+m',\,n+n'}(g)}\;\braket{jn,j'n'}{k\,(n\!+\!n')}.\label{eq:pw4}
\end{align}
\noindent This is Eq.~\eqref{eq:multiplySU2gaugemodeswithCG}. In the label notation of Eq.~\eqref{eq:SU2gaugefield}, this becomes the $\odot$ product of Eq.~\eqref{eq:multiplySU2gaugelabels}.

\end{document}